\documentclass[twocolumn,showpacs,preprintnumbers,amsmath,amssymb,superscriptaddress]{revtex4-2}

\usepackage{graphicx}
\usepackage{dcolumn}
\usepackage{bm}
\usepackage{tabularx}
\usepackage{makecell}
\usepackage{braket}
\usepackage{lipsum}
\usepackage{ulem}
\usepackage{amsmath,amssymb}
\newcolumntype{M}{>{\centering\arraybackslash}m{1.85cm}}
\usepackage[export]{adjustbox}
\usepackage{longtable}
\usepackage{float}
\usepackage{xcolor}
\usepackage{color}   
\usepackage{hyperref}
\hypersetup{
	colorlinks=true, 
	linktoc=all,     
	linkcolor=blue,  
}

\makeatletter
\newcommand{\colorcaption}[2][]{%
	\begingroup%
	\renewcommand{\@caption@fignum@sep}{ (Color online). }%
	\caption[#1]{#2}%
	\endgroup%
}
\makeatother
\newcommand{\orcid}[1]{\href{https://orcid.org/#1}{\hskip2pt\includegraphics[width=9pt]{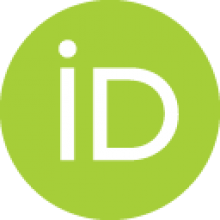}}}
\begin{document}

 	\title{Weak interaction rates of $sd$-shell Urca pairs in stellar environment using a shell-model approach}

	\author{Shweta Sharma\orcid{0000-0003-4146-383X}}
	\address{Department of Physics, Indian Institute of Technology Roorkee, Roorkee 247667, India}
	\author{Praveen C. Srivastava\orcid{0000-0001-8719-1548}}
	\email{Corresponding author: praveen.srivastava@ph.iitr.ac.in}
	\address{Department of Physics, Indian Institute of Technology Roorkee, Roorkee 247667, India}

	\author{Toshio Suzuki\orcid{0000-0002-5500-539X}}
	\address{Department of Physics, College of Humanities and Sciences, Nihon University, Sakurajosui 3, Setagaya-ku, Tokyo 156-8550, Japan}
	\address{NAT Research Center, NAT Corporation, 3129-45 Hibara Muramatsu, Tokai-mura, Naka-gun, Ibaraki 319-1112, Japan}
	\address{School of Physics, Beihang University, 37 Xueyuan Road, Haidian District, Beijing 100191, People's Republic of China}

	\date{\hfill \today}
	\begin{abstract}
	In this work, the weak interaction rates of $sd$-shell nuclei in the hot and dense stellar environment are calculated using the shell model. The {\it ab initio} effective interactions are employed for the calculation of Gamow-Teller strengths and weak rates in addition to the phenomenological interaction. The weak rates are evaluated in a fine grid of density and temperature. The electron capture and $\beta^-$-decay rates of Urca pairs of nuclei with $A = 23, 25, 29, 31$ and 33 are evaluated as a function of $\log_{10} \rho Y_e$ and the corresponding Urca density is obtained. The screening effects are also included for the evaluation of these stellar weak rates. In addition, the intrinsic cooling strength is evaluated for $A=29,31$ and 33 Urca pairs in neutron star crusts. The stellar weak rates, along with the neutrino energy emitted and the gamma heat production, are tabulated for $A = 29-39$ nuclei.

	\end{abstract}

	\maketitle

	\section{Introduction}
The study of the weak interaction rates in stellar environments is important for many astrophysical applications like comprehending the cooling stages of the stellar evolution, neutron star crusts, and oceans, and in presupernova stellar collapse \cite{burbidge1957, langanke2014}. During a supernova, the star core experiences a significant increase in density, and the Fermi energy of the electrons increases as the electron gas becomes degenerate \cite{langanke2003}. This causes the capture of degenerate relativistic electrons by the nucleus because the Fermi energy of the degenerate electrons exceeds the negative $Q$ value for the capture of electrons by the nuclei \cite{takahara1989}. This phenomenon is referred to as the electron capture (e-capture) process. This process leads to a decrease in the degenerate electron pressure within the stellar core and cooling of the stellar core, along with a decrease in entropy as the neutrinos escape from the star \cite{suzuki2011, pinedo2014, pinedo2022, langanke2015}. Further, the Urca process is a major phenomenon responsible for the cooling of the stars. In the Urca process, the electron capture and its inverse process $\beta^-$ decay occur simultaneously, thereby emitting neutrinos and antineutrinos from the stellar core.

For the first time, electron capture rates were studied by Mazurek {\it et al.} \cite{mazurek1974} and Duke {\it et al.} \cite{duke1971}. Fuller, Fowler, and Newman studied the weak interaction rates for a wide range of nuclei $21 \le A \le 60$ using an independent particle shell model and also gave the general formalism for these processes \cite{fuller1980,fuller1982,fuller1982s,fuller1985}. Further, Oda {\it et al.} \cite{oda1994} studied these weak interaction rates for sd-shell nuclei using nuclear shell model. The authors also used experimental data of $B$(GT)  wherever available and showed weak rates in the grids of different temperatures and densities. Following this, various studies were carried out to calculate the weak interaction rates of $pf$-shell nuclei \cite{caurier1999, langanke2000, sarriguren2016, cole2012}. 

In addition, the importance of forbidden transitions in
electron capture rates during stellar core collapse was investigated and recognized in Refs. \cite{CW1984,juo10}. The importance of the second-forbidden transition, $^{20}$Ne($0^+$) $\rightarrow$ $^{20}$F($2^+$), was first pointed out in Ref. \cite{pinedo2014}, and the e-capture rates were
calculated by the multipole expansion method \cite{suzuki2019}. The weak rates were successfully evaluated with the conserved-vector current constraints in Refs. \cite{kirsebom2019a,kirsebom2019b,suzuki2022}.
Electron capture rates in the second-forbidden transitions, $^{24}$Na (4$_{\rm g.s.}^{+}$) $\rightarrow$ $^{24}$Ne (2$_{1}^{+}$) and $^{27}$Al (5/2$_{\rm g.s.}^{+}$) $\rightarrow$ $^{27}$Mg (1/2$_{\rm g.s.}^{+}$), were also evaluated in oxygen-neon stellar cores \cite{stromberg2022}.   

The evaluation of Gamow-Teller (GT) strengths is crucial
for the study of weak interaction rates. There are two types of GT strengths, i.e., GT$^+$ and GT$^-$ strength, dependent on the type of weak interaction process. The experimental GT strengths can be obtained through charge-exchange reactions and decays, but the energy resolution and ranges are limited. Therefore, we need theoretical models that can accurately estimate these GT strengths. The interacting shell model with
an appropriate effective interaction is considered the most efficient model for the calculation of these GT strengths. The phenomenological USDB effective interaction \cite{brown2006} has demonstrated its efficacy in describing the nuclear structure properties of $sd$-shell nuclei. The USDB interaction utilizes a
renormalized G matrix that incorporates linear combinations of two-body matrix components calibrated accurately to a collection of experimental data for $sd$-shell nuclei. This effective interaction was used for the evaluation of weak interaction rates of $sd$-shell nuclei for $A=17-28$ \cite{suzuki2016}. We intend to extend this study further for $sd$-shell nuclei with $A=29-39$.

Further, over the past few years, the {\it ab initio} effective interaction has gained a lot of interest in the study of nuclear structure. In the {\it ab initio} effective interaction, the effective two-body interaction term is obtained by performing {\it ab initio} no core shell-model calculations and the Okubo-Lee-Suzuki transformation. This results in the separation of the effective Hamiltonian into inert core, one-body, and two-body components. One such {\it ab initio} interaction is the Daejeon 16A
(DJ16A) effective interaction \cite{smirnova2019, shirokov2016}, which is the monopole corrected version of Daejeon 16 (DJ16) constructed for $sd$-shell nuclei. We have used this effective interaction for the evaluation of weak interaction rates in our work in order to observe the impact of microscopic interactions on weak interaction rates for $sd$-shell nuclei.

Another effective interaction used in our work is obtained from chiral effective field theory (CEFT), and derived by Huth {\it et al.} \cite{huth2018}. In this interaction, the Weinberg power counting is used for the expansion of the $(Q/\Lambda_b)^{\nu}$ term with order $\nu \ge 0$, where $Q$ is the low momentum scale and $\Lambda_b$ is the breakdown scale of the EFT. In this approach, the Hamiltonian for the sd shell is constructed with two low-energy constants (LECs) at leading order along with seven new LECs at next-to-leading order. Further, the LECs of CEFT operators are fitted directly to 441 ground-state (g.s.) and excited-state energies. These successful microscopic interactions inspire us to employ them for the study of electron capture rates of $sd$-shell nuclei.

In addition, we have also used SDPF-MU realistic interaction for the calculation of GT strengths for Urca pair nuclei ($^{31}$Al,$^{31}$Mg), ($^{31}$Si,$^{31}$P) and ($^{33}$Al,$^{33}$Mg) in order to take account of higher $pf$-shell orbitals in the configuration space. This interaction was constructed by Utsuno {\it et al.} \cite{utsuno2012} with the proton-neutron cross-shell interaction provided by the V$_{\rm MU}$ universal interaction \cite{otsuka2010}. An effective interaction in the $sd$-$pf$ shell obtained by the extended Kuo-Krenciglowa (EKK) method \cite{tsunoda2017,tsunoda2020}, which proved to be successful in describing the nuclei in the island of inversion, is also used for the study of the ($^{31}$Al, $^{31}$Mg) pair. We investigate the differences and similarities in the GT strengths and weak rates among the {\it ab initio} and phenomenological interactions. We are especially interested in whether Urca densities can also be assigned for {\it ab initio} effective interactions.

Section \ref{beta} is devoted to the brief formalism behind the weak interaction rates in the stellar environment. Section \ref{result} shows the results for the GT strengths and the weak rates in stellar environments, together with a comprehensive discussion. An application to Urca cooling in neutron star crusts with the use of the present weak rates is also carried out in this section. A summary and conclusion are given in Sec. \ref{Conclusion}.

\section{ Formalism for stellar weak interaction rates} \label{beta}

\subsection{Electron chemical potential ($\mu_e$)}

The electron chemical potential is determined by the density expression as
\begin{equation}
\rho Y_e=\frac{1}{\pi^2 N_A}(\frac{m_e c}{\hbar})^3\int_0^\infty (S_e-S_p)p^2 dp,
\end{equation}
where $\frac{(m_ec)^3}{\pi^2\hbar^3 N_A}=2.9218 \times 10^6 \text{g cm}^{-3}$ and the terms $\rho$ and $Y_e$ are the density and the baryon ratio, respectively. The term $S_e$ stands for the Fermi-Dirac distribution of electrons, and $S_p$ is subtracted in the equation to cancel out the effect of positrons. $S_e$ is defined as
\begin{equation}
S_e=\frac{1}{exp(\frac{E_e- \mu_e}{kT})+1},
\end{equation}
where $E_e$ is the electron energy and $\mu_e$ is the electron chemical potential. In addition, $\mu_e$ is replaced by $\mu_p$ in the expression for $S_p$ where $\mu_p=-\mu_e$.\\

\subsection{Weak interaction rates}

The weak interaction rates are evaluated using the values of the chemical potential evaluated above. The formulation used for this calculation is given here. The weak interaction rate in the stellar environment can be evaluated as\\

\begin{equation}
\lambda=\frac{\rm{ln}2}{K}\sum_i W_i\sum_j (B_{ij}({\rm GT})+B_{ij}({\rm F}))\phi_{ij}^{\alpha},
\end{equation}
where $i$ and $j$ stand for initial and final states, respectively. The constant $K = 6146 \pm 6$s \cite{haxton1995} is determined from superallowed Fermi transitions. The term $W_i$ is the statistical weight function, and $B_{ij}({\rm GT})$ and $B_{ij}({\rm F})$ \cite{brown1985, saxena2018} are GT and Fermi-reduced transition probabilities, respectively. The term $\alpha$ = ec, $\beta^+$, pc, $\beta^-$ for electron capture, $\beta^+$ decay, positron capture , and $\beta^-$ decay, respectively. Thus, the term $\phi_{ij}^{\alpha}$ \cite{langanke2001} is given by

\begin{equation}\label{eq1}
\phi_{ij}^{ec}=\int_{w_{min}}^{\infty}wp(Q_{ij}+w)^2F(Z,w)S_e(w)dw,
\end{equation}

\begin{equation}
\phi_{ij}^{\beta^+}=\int_{1}^{Q_{ij}}wp(Q_{ij}-w)^2F(-Z+1,w)[1-S_p(w)]dw,
\end{equation}

\begin{equation}\label{eq1}
\phi_{ij}^{pc}=\int_{w_{min}}^{\infty}wp(Q_{ij}+w)^2F(-Z,w)S_p(w)dw,
\end{equation}

\begin{equation}
\phi_{ij}^{\beta^-}=\int_{1}^{Q_{ij}}wp(Q_{ij}-w)^2F(Z+1,w)[1-S_e(w)]dw.
\end{equation}

Here, $w$ and $p=\sqrt{w^2-1}$ are electron energy and momentum, respectively. $F(Z,w)$ is the Fermi function, and the quantity $S_e(w)$ and $S_p(w)$ is the Fermi-Dirac distribution for electrons and positrons, respectively. The integration lower limit $w_{min}=1$ if $Q_{ij} \geq -1$ and $w_{min}=|Q_{ij}|$ if $Q_{ij}<-1$ where $Q_{ij}$ is the $Q$ value which is given as

\begin{equation}
Q_{ij}=(Q_{\rm g.s.}+E_i-E_f)/m_e c^2,
\end{equation}
where the $E_i$ and $E_f$ are the excitation energies of the initial and final states, respectively. The term $Q_{\rm g.s.}$ is the $Q$-value between the ground states which is defined as
\begin{equation}
Q_{\rm g.s.}=M_p c^2-M_d c^2,
\end{equation}
where $M_p$ and $M_d$ are the nuclear masses of parent and daughter nuclei, respectively. Further, the statistical weight function is given by,
\begin{equation}
W_i=(2 J_i+1)e^{-E_i/kT}/\sum_i (2 J_i+1)e^{-E_i/kT},
\end{equation}
where $J_i$ is the angular momentum of the initial state.
The quantities  $B_{ij}({\rm GT})$ and $B_{ij}({\rm F})$ are given by
\begin{equation}
B_{ij}({\rm GT})=(g_A/g_V)^2 \frac{1}{2J_i+1}|<f||\sum_k\sigma^k t_+^k||i>|^2,
\end{equation}

\begin{equation}
B_{ij}({\rm F})= \frac{1}{2J_i+1}|<f||\sum_k t_+^k||i>|^2.
\end{equation}

Here, $g_A$ and $g_V$ are the axial vector and vector coupling constants. $g_A^{eff}=qg_A=0.77\times 1.27=0.98$ and $g_V=1.00$ were taken for calculations. The $B_{ij}$(GT) and $B_{ij}$(F) values have been taken from the shell-model calculation for different transitions using the \small{KSHELL} and \small{NUSHELLX} shell-model codes \cite{kshell, nushellx}.

Further, the Fermi function $F(Z,w)$ is defined as
\begin{equation}
F(Z,w)=(2)(1+\gamma)e^{\pi{y}}\left(\frac{2pR}{\hbar}\right)^{2(\gamma-1)}\left(\frac{|\Gamma(\gamma+iy)|}{\Gamma(1+2\gamma)}\right)^2.
\end{equation}
Here, $p=\sqrt{w^2-1}$ and $\gamma$ and $y$ are the auxiliary quantities, which are given as
$\gamma =\sqrt{1-(\alpha{Z})^2}$ and $y=(\alpha{Zw}/pc)$, where $\alpha=1/137$ is the fine structure constant.

\subsection{Screening effects}
The screening effects of the degenerate electrons in a stellar environment affect the weak interaction rates. The first screening effect changes the threshold energy because of the change in the chemical potential of the ions. The $Q_{ij}$ is modified as
\begin{equation}
Q_{ij}=Q_{ij}+\Delta Q_C,
\end{equation}
where $\Delta Q_C$ is defined as

\begin{equation}\label{delq}
\Delta Q_C=\mu_C(Z_p)-\mu_C(Z_d),
\end{equation}
where $\mu_C(Z)$ is the Coulomb chemical potential,
and $Z_p$ and $Z_d$ are the nuclear charges of parent and daughter nuclei, respectively.
$\Delta Q_C$ is positive (negative) in $\beta^{-}$-decay (electron capture) process.
Detailed expressions for  $\mu_C(Z)$ can be found in Refs. \cite{ichimaru93,juo10}.\\

Second is the screening effects of electrons where chemical potential is changed by
\begin{equation}
\mu_e \rightarrow \mu_e-V_s(0),
\end{equation}
where $V_s(0)$ is Coulomb potential at origin, which at a distance $r$ is given by
\begin{equation}
V_s(r)=Ze^2(2k_F)J(r),
\end{equation}
where $J(r)$ is given by
\begin{equation}
J(r)=\frac{1}{2k_Fr}\big[ 1-\frac{2}{\pi}\int\frac{sin(2k_Fqr)}{q^2\epsilon(q,0)} dq \big],
\end{equation}
and the analytic fitting formula is used for the evaluation of $J(r)$ as given in Ref. \cite{itoh2002}.

\subsection{ Neutrino-energy-loss rate ($\xi$)}
The energy loss by neutrino emission is given by
\begin{equation}
\xi=\frac{{\rm ln2} (m_ec^2)}{6146(s)}\sum_i W_i\sum_j [B_{ij}({\rm GT})+B_{ij}({\rm F})]\psi_{ij}^{\alpha},
\end{equation}
where $\psi_{ij}^{\alpha}$ is expressed as
\begin{equation}
\psi_{ij}^{ec}=\int_{w_{min}}^{\infty}wp(Q_{ij}+w)^3F(Z,w)S_e(w)dw,
\end{equation}

\begin{equation}
\psi_{ij}^{\beta^+}=\int_{1}^{Q_{ij}}wp(Q_{ij}-w)^3F(-Z+1,w)[1-S_p(w)]dw.
\end{equation}

\begin{equation}
\psi_{ij}^{pc}=\int_{w_{min}}^{\infty}wp(Q_{ij}+w)^3F(-Z,w)S_p(w)dw,
\end{equation}

\begin{equation}
\psi_{ij}^{\beta^-}=\int_{1}^{Q_{ij}}wp(Q_{ij}-w)^3F(Z+1,w)[1-S_e(w)]dw.
\end{equation}

Now, the average energy of the emitted neutrino is given by
\begin{equation}
<E_{\nu}>=\frac{\xi}{\lambda}.
\end{equation}

\subsection{ $\gamma$-ray heating rate ($\eta$):}
The heating rate by $\gamma$ emission is given by
\begin{equation}
\eta=\frac{{\rm ln2} (m_ec^2)}{6146(s)}\sum_i W_i\sum_j [B_{ij}({\rm GT})+B_{ij}({\rm F})]\Gamma_{ij}^{\alpha},
\end{equation}
where $\Gamma_{ij}^{\alpha}$ is expressed as
\begin{equation}
\Gamma_{ij}^{ec}=\int_{w_{min}}^{\infty}wp(Q_{ij}+w)^2 E_f/m_ec^2F(Z,w)S_e(w)dw,
\end{equation}

\begin{equation}
\Gamma_{ij}^{\beta^+}=\int_{1}^{Q_{ij}}wp(Q_{ij}-w)^2E_f/m_ec^2F(-Z+1,w)[1-S_p(w)]dw.
\end{equation}

\begin{equation}
\Gamma_{ij}^{pc}=\int_{w_{min}}^{\infty}wp(Q_{ij}+w)^2E_f/m_ec^2F(-Z,w)S_p(w)dw,
\end{equation}

\begin{equation}
\Gamma_{ij}^{\beta^-}=\int_{1}^{Q_{ij}}wp(Q_{ij}-w)^2E_f/m_ec^2F(Z+1,w)[1-S_e(w)]dw.
\end{equation}

Now, the average energy of the emitted $\gamma$ is given by
\begin{equation}
<E_{\gamma}>=\frac{\eta}{\lambda}.
\end{equation}

\begin{figure*}\label{bgt_23na}
 \includegraphics[width=82mm]{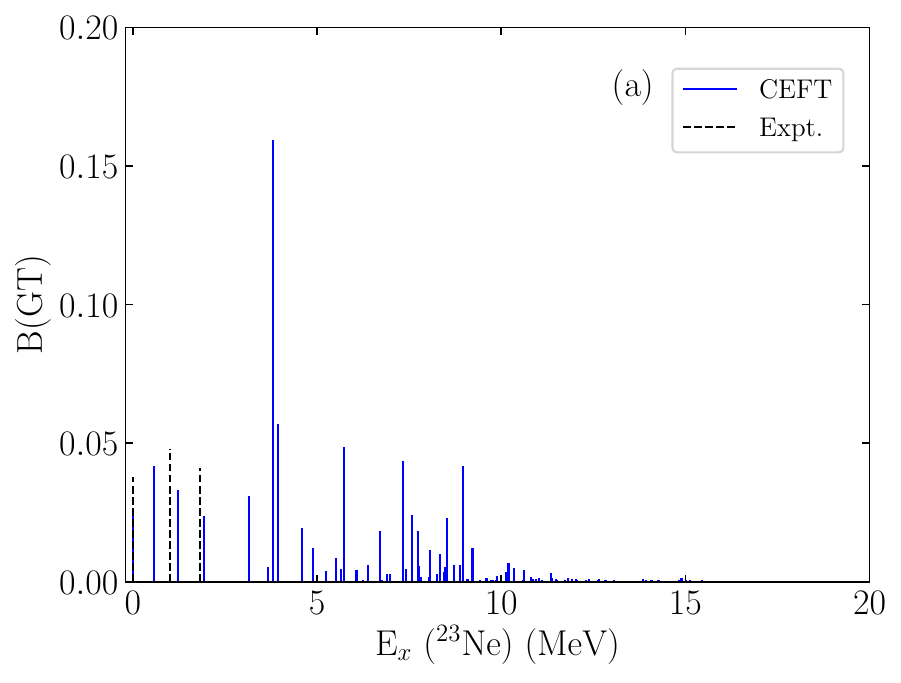}
\includegraphics[width=82mm]{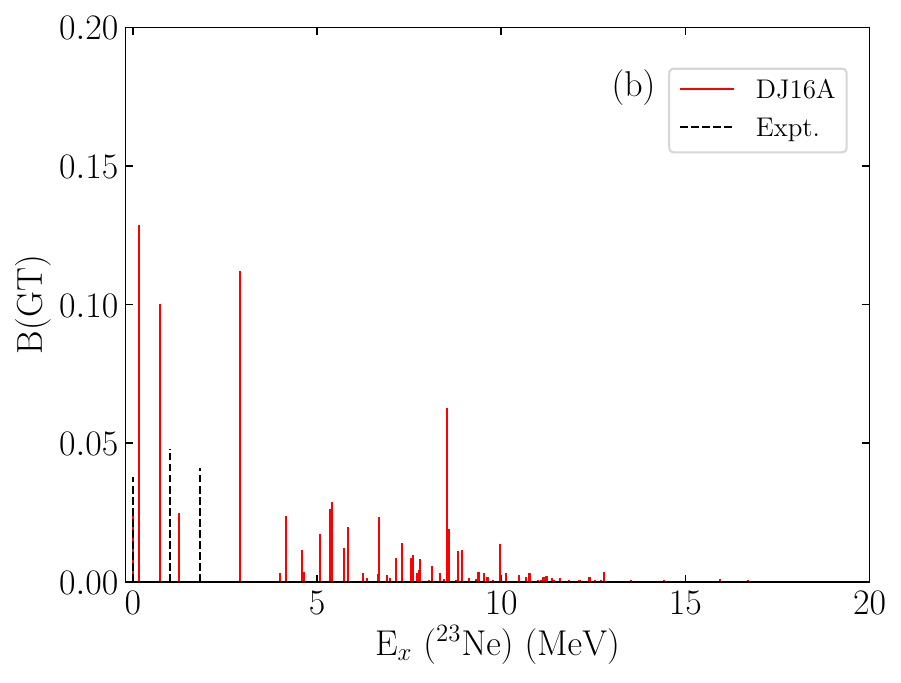} \includegraphics[width=82mm]{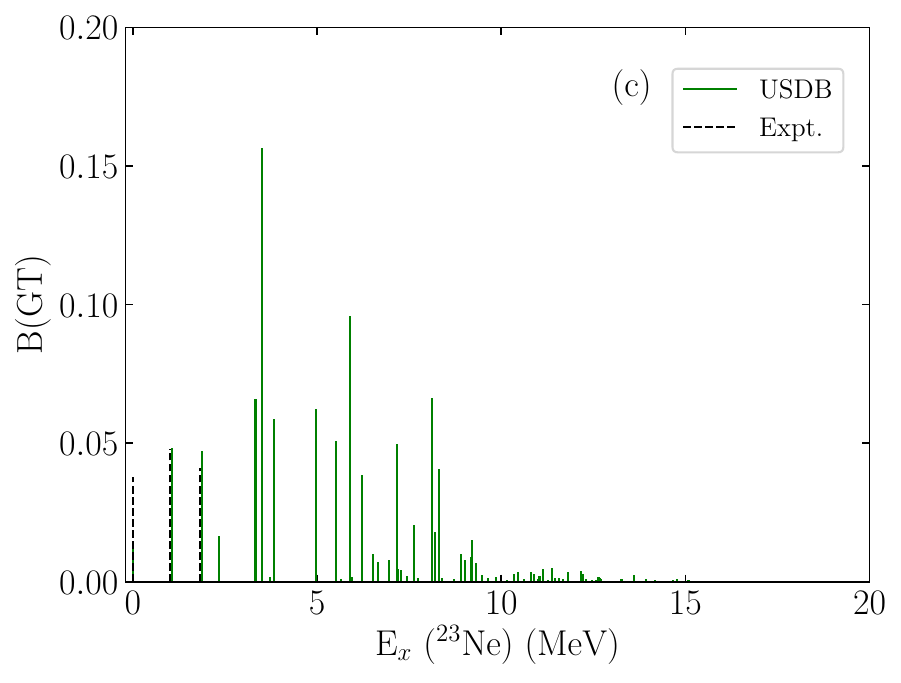}
\includegraphics[width=82mm]{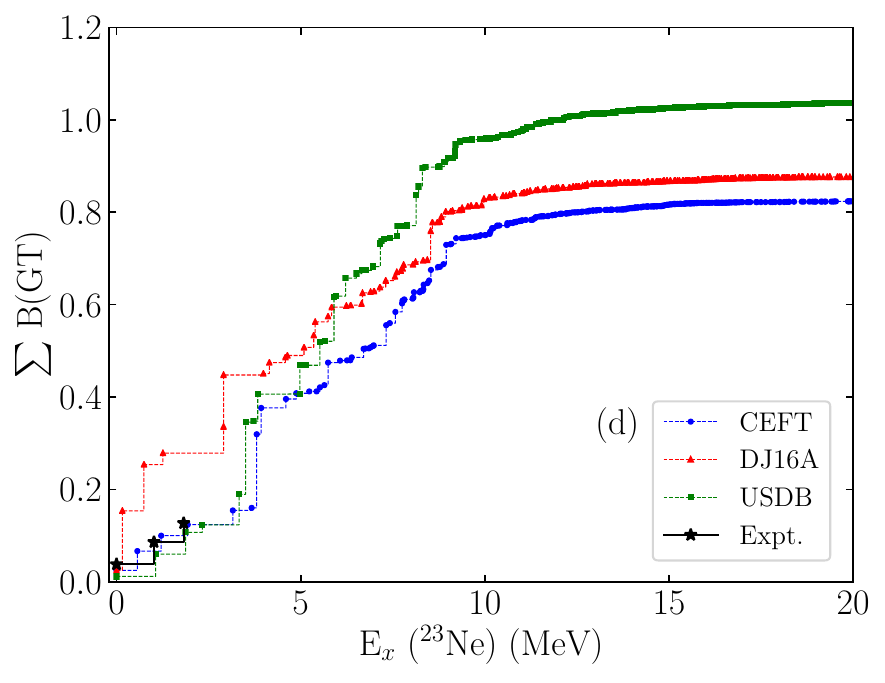}
	\caption{$B$(GT) and $\sum B$(GT) values from $3/2^+_{\rm g.s.}$ of $^{23}$Na to various excited states of $^{23}$Ne calculated  using CEFT, DJ16A, and  USDB  effective interactions and comparison with the experimental data.\label{bgt_23na}}
\end{figure*}

 \section{Results and Discussion} \label{result}
 \subsection{Gamow-Teller strengths and weak rates for Urca pairs with $A$ = 23 and 25}

The electron capture and $\beta^-$ decay rates for nuclei pairs with $A = 23$ and 25 are of significant importance, as these pairs participate in Urca processes within the O-Ne-Mg core of stars with eight to ten solar masses \cite{Toki2013}. These nuclear pairs play a crucial role in the cooling mechanisms of the O-Ne-Mg core. Thus, the weak interaction rates for ($^{23}$Na, $^{23}$Ne), ($^{25}$Mg, $^{25}$Na), ($^{25}$Na, $^{25}$Ne) are evaluated at higher temperature and density regimes. Since the $B$(GT) values are an important input for determining weak interaction rates, we have compared these $B$(GT) values computed through various effective interactions and available experimental data. Here, we have evaluated GT strengths from the ground and excited states of the parent nuclei up to an excitation energy of 2 MeV, to modestly sufficient excited states of the corresponding daughter nuclei.  

In Fig. \ref{bgt_23na}, the $B$(GT) and cumulative $\sum B$(GT) values from the g.s. $3/2^+_{\rm g.s.}$ of $^{23}$Na to both ground and various excited states of $^{23}$Ne are shown. These $B$(GT) values are plotted as a function of the excitation energy of the daughter nucleus, with three different effective interactions CEFT, DJ16A, and USDB used for the evaluation of GT strengths. A detailed comparison is made between the $B$(GT) values obtained through these interactions and the available experimental data. The corresponding experimental data are taken from Ref. \cite{siebels1995} where measurements for GT strengths are performed via $^{23}$Na (n,p)$^{23}$Ne reaction at the TRIUMF charge exchange facility. The measured B$^+$(GT) values
corresponding to $5/2^+_{\rm g.s.}$, $1/2^+$(1.017 MeV), $3/2^+$ (1.823 MeV) and $5/2^+$ (2.315 MeV) states of $^{23}$Ne are 0.038$\pm$0.006, 0.048$\pm$0.007, 0.041$\pm$0.008 and $\le$0.021, respectively. The experimental $B$(GT) values corresponding to these excited states were not considered in Ref. \cite{suzuki2016}. The GT strengths calculated from shell-model calculations are scaled by a quenching factor of 0.77. The $B$(GT) values computed from CEFT and USDB effective interactions are in reasonable agreement with the experimental data, but DJ16A interaction tends to overestimate the $B$(GT) values at lower excitation
energies of daughter nuclei. Furthermore, the GT strengths
obtained from USDB interaction are more dominant in the
5-10 MeV excitation energy range compared to other interactions. A prominent peak is seen around 3-4 MeV in all
three interactions, which is due to the GT transition from the $3/2^+_{\rm g.s.}$ of $^{23}$Na to the excited state $1/2^+_2$ of $^{23}$Ne. These peaks can be seen at 3.807, 2.906, and 3.508 MeV excitation energies for CEFT, DJ16A, and USDB interactions, respectively. The $B$(GT) strengths diminish after approximately 11 MeV. Figure \ref{bgt_23na}(d) shows the cumulative sum of GT strengths from all three interactions and experimental data. From this plot, it is evident that the cumulative sums of the $B$(GT) values saturate after around 11 MeV excitation energy.  While the DJ16A interaction dominates in the low energy region below 5 MeV, the USDB interaction gives larger GT strength peaks at higher energies, resulting in the highest cumulative sum of $B$(GT) values of the USDB interaction at higher energies as compared to other interactions.

 In the same way, the $B$(GT) and $\sum B$(GT) values from the g.s. $5/2^+_{\rm g.s.}$ of $^{25}$Mg to the ground and various excited states of $^{25}$Na are evaluated with  the three shell-model interactions CEFT, DJ16A and USDB. The $B$(GT) obtained from all three interactions agree very well with the experimental $B$(GT) value, which is available only for the g.s. to g.s. transition. Similar to the case of $^{23}$Na, the cumulative sum of $B$(GT) values is largest for DJ16A interaction up to an excitation energy of 5 MeV, while for energies beyond 5 MeV, the cumulative sum of the $B$(GT) values from the USDB interaction surpasses that from DJ16A.

 Next, we briefly discuss the GT strength in $^{25}$Na.
The $B$(GT) and $\sum B$(GT) values from the g.s. $5/2^+_{\rm g.s.}$ of $^{25}$Na to the ground and various excited states of $^{25}$Ne are evaluated by the shell model using the three interactions. Here, the transition from the 5/2$_{\rm g.s.}^{+}$ state in $^{25}$Na to the 1/2$_{\rm g.s.}^{+}$ state in $^{25}$Ne is second nonunique forbidden. However, DJ16A effective interaction predicts $3/2^+$ as the g.s. for $^{25}$Na isotope which makes the g.s. to g.s. transition allowed for DJ16A effective interaction. A strong peak is 
found around 2 MeV, corresponding to the transition from the $5/2^+_{\rm g.s.}$ in $^{25}$Na to the $3/2^+_1$ in $^{25}$Ne. All three interactions exhibit a similar trend in the spread of the GT strengths, with these strengths starting to diminish after approximately 10 MeV excitation energy. The USDB interaction gives the highest cumulative value, indicating a stronger overall contribution to the GT strengths compared to the other interactions.
 Note that there exists a 3/2$^{+}$ state at very low excitation energy of 0.0895 MeV in $^{25}$Na, and the GT transitions from this state give dominant contributions to the weak rates in stellar environments.

\begin{figure}[h]\label{rate_23}
 \includegraphics[width=77mm]{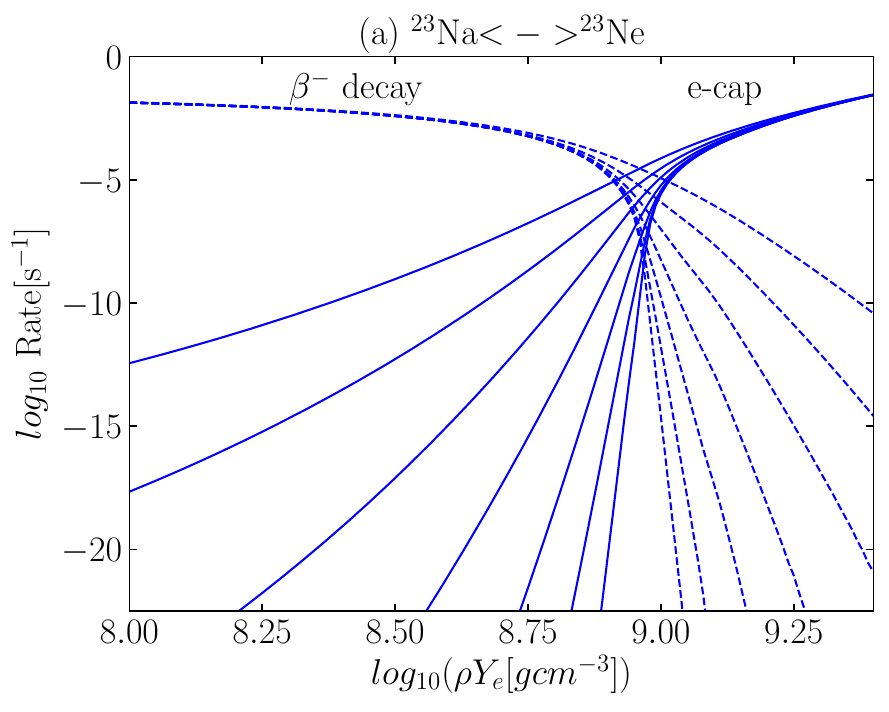}
 \includegraphics[width=77mm]{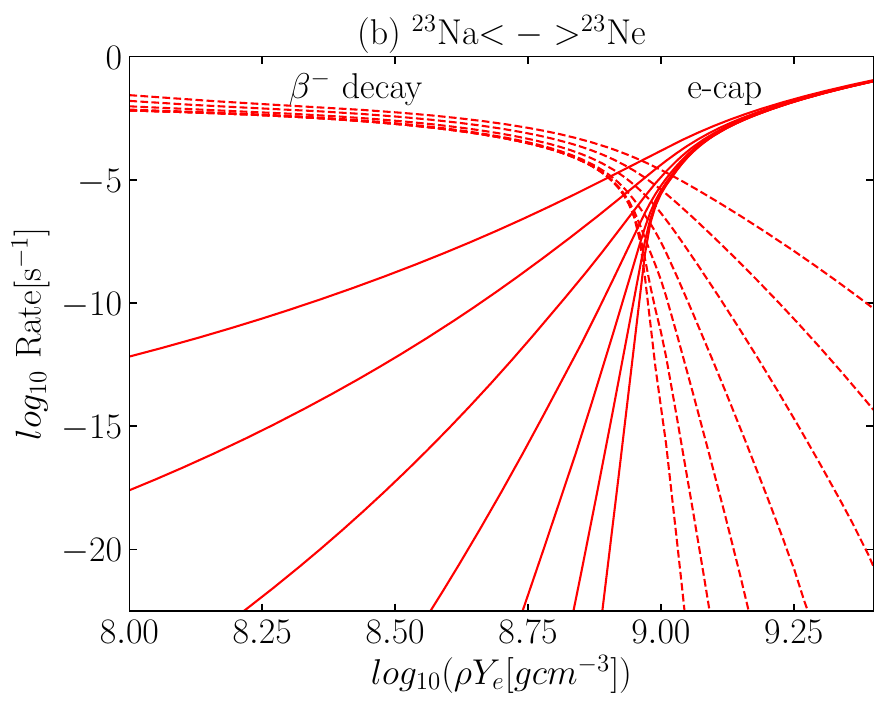}
 \includegraphics[width=77mm]{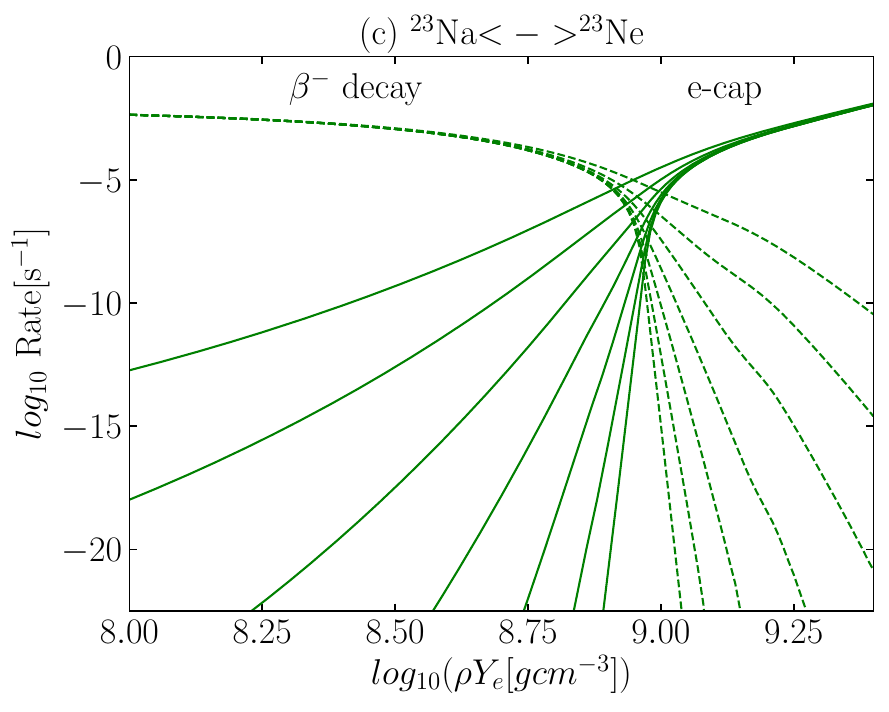}
	\caption{Electron capture and $\beta^-$ decay rates for ($^{23}$Na, $^{23}$Ne) at various temperatures $\log_{10}T=8.0-9.2$ in the interval of 0.2 with Coulomb effects using (a) CEFT, (b) DJ16A, and  (c) USDB effective interactions.\label{rate_23}}
\end{figure}

\begin{figure}[h]\label{rate_23_expt}
 \includegraphics[width=77mm]{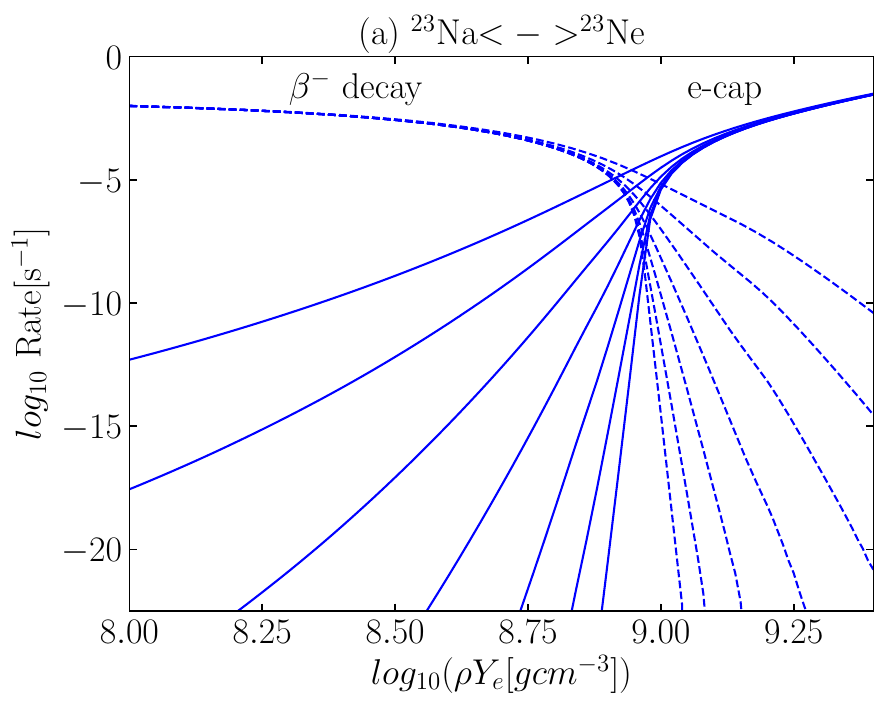}
 \includegraphics[width=77mm]{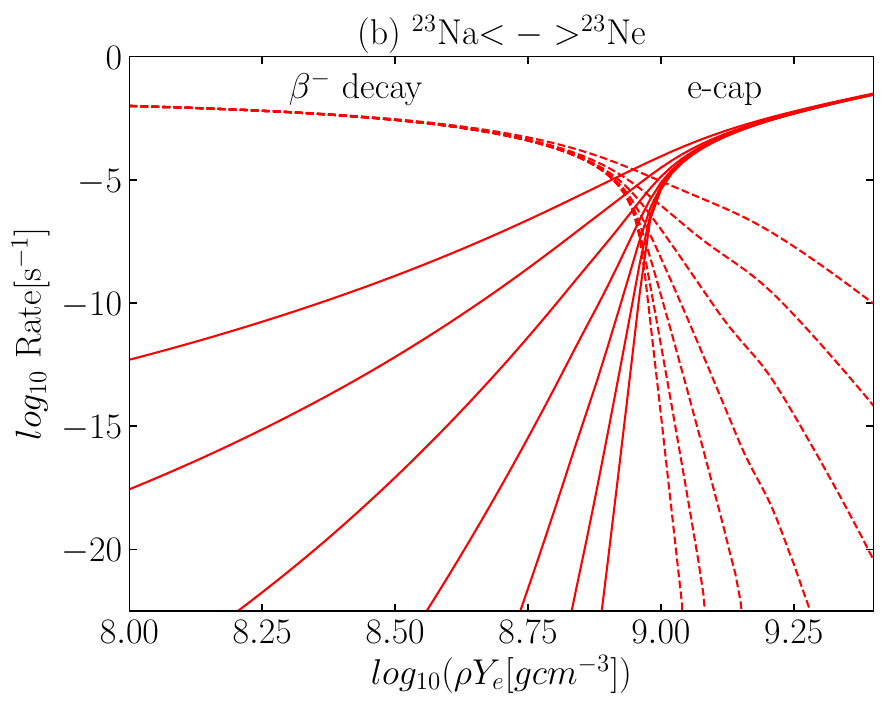}
 \includegraphics[width=77mm]{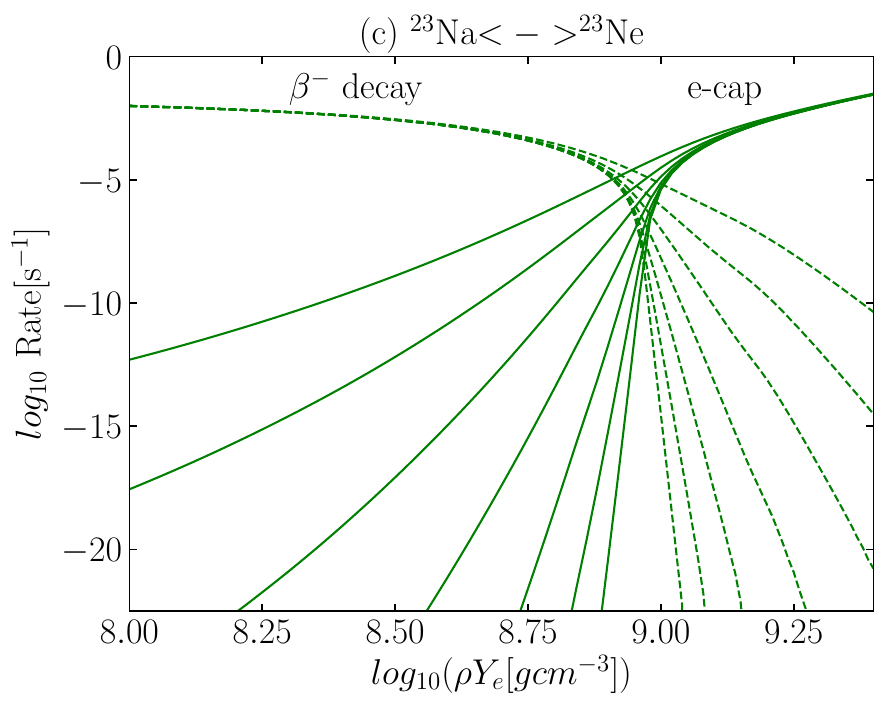}
	\caption{Electron capture and $\beta^-$ decay rates for ($^{23}$Na, $^{23}$Ne) at various temperatures $\log_{10}T=8.0-9.2$ in the interval of 0.2 with Coulomb effects using experimental data along with (a) CEFT, (b) DJ16A, and  (c) USDB effective interactions.\label{rate_23_expt}}
\end{figure}

 \begin{figure}[h]\label{rate_23na}
 \includegraphics[width=70mm]{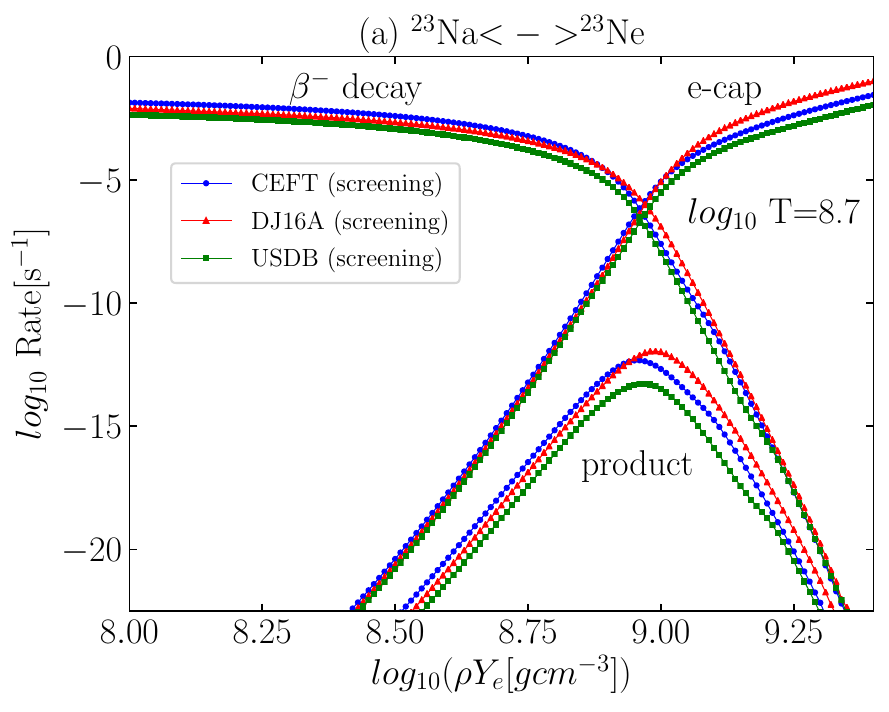}
 \includegraphics[width=70mm]{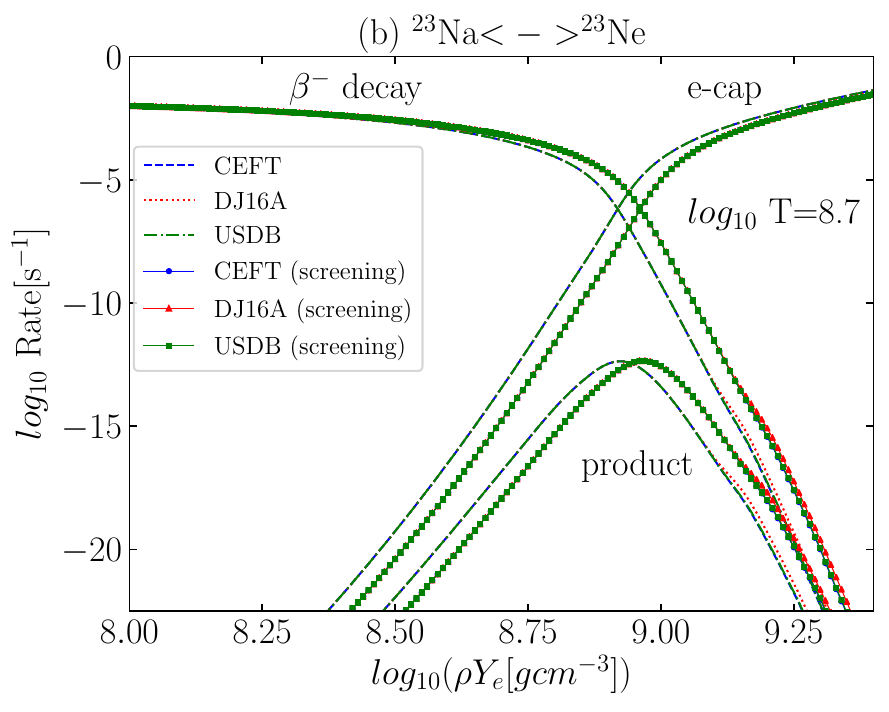}
 \includegraphics[width=70mm]{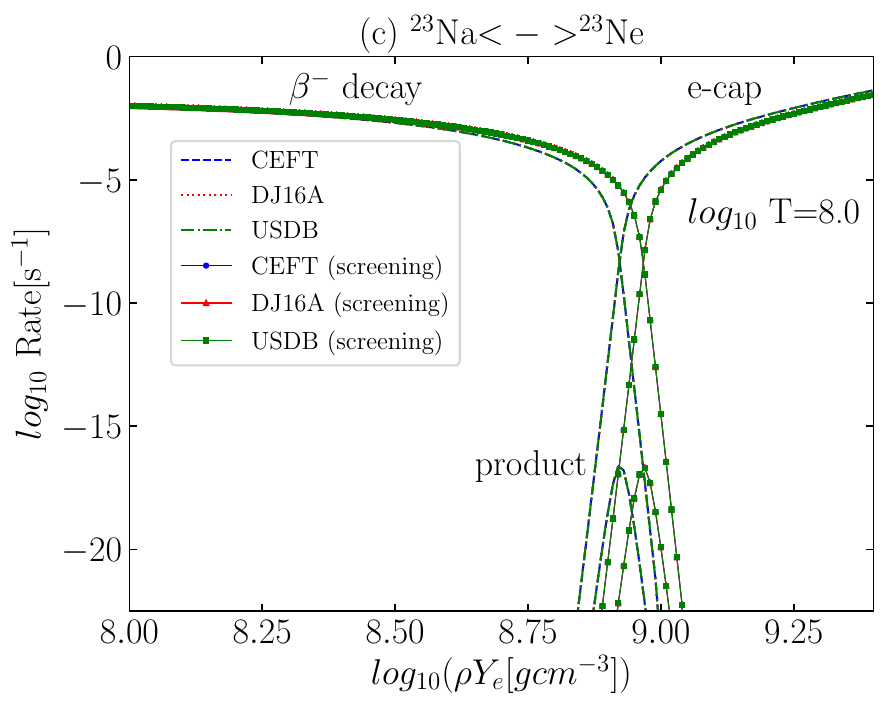}
	\caption{Electron capture and $\beta^-$ decay rates for ($^{23}$Na, $^{23}$Ne) pair at $\log_{10}T=8.7,8.0$ with and without Coulomb effects using CEFT, DJ16A and USDB effective interactions.
   Available experimental data for the excitation energies and $B$(GT) values are included in (b) and (c), while they are not included in (a).
\label{rate_23na}}
\end{figure}

\begin{figure}[h]\label{rate_25}
 \includegraphics[width=77mm]{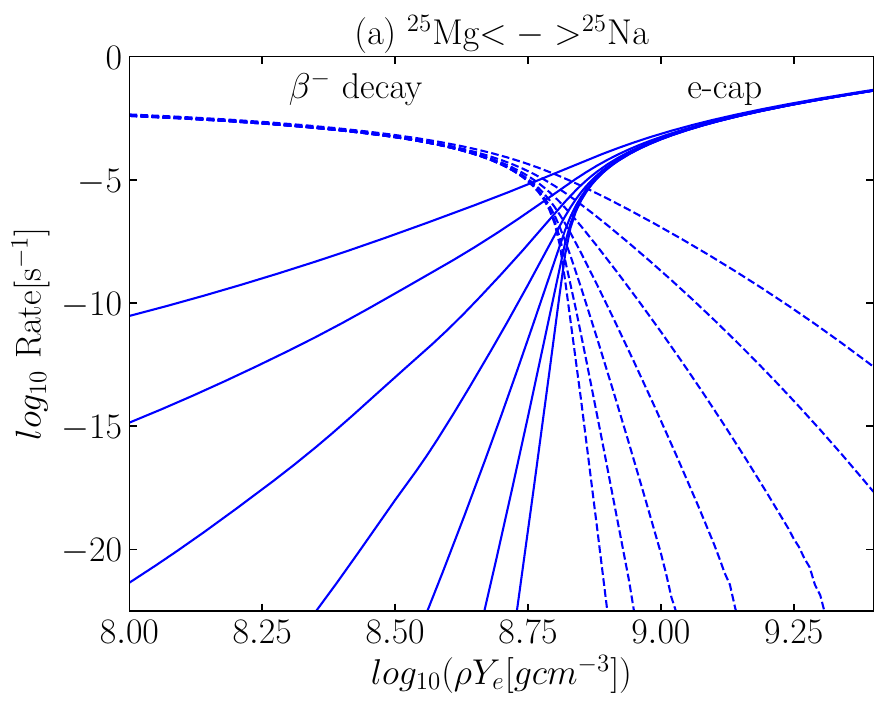}
 \includegraphics[width=77mm]{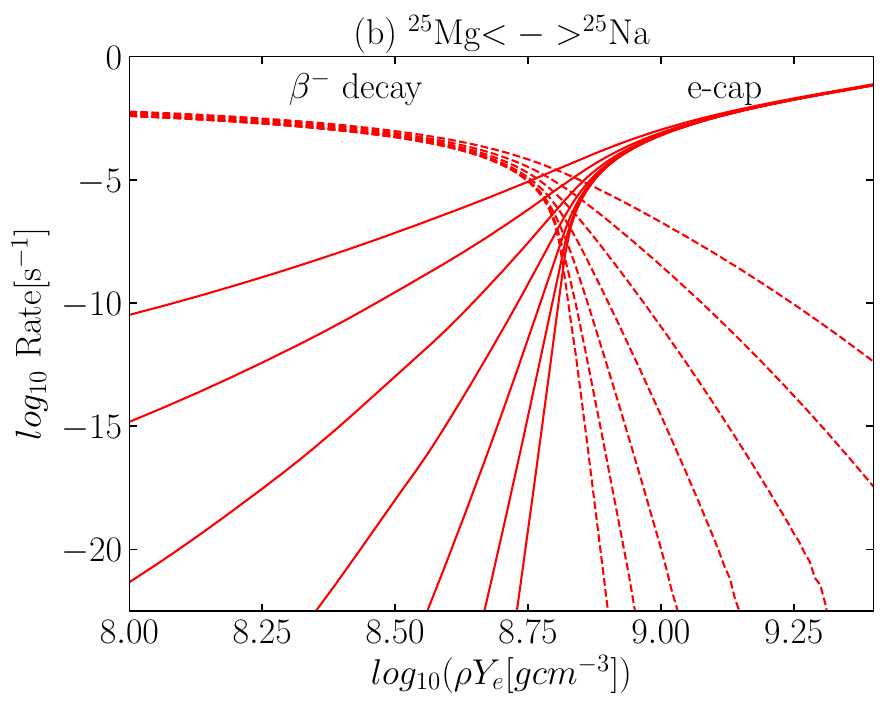}
 \includegraphics[width=77mm]{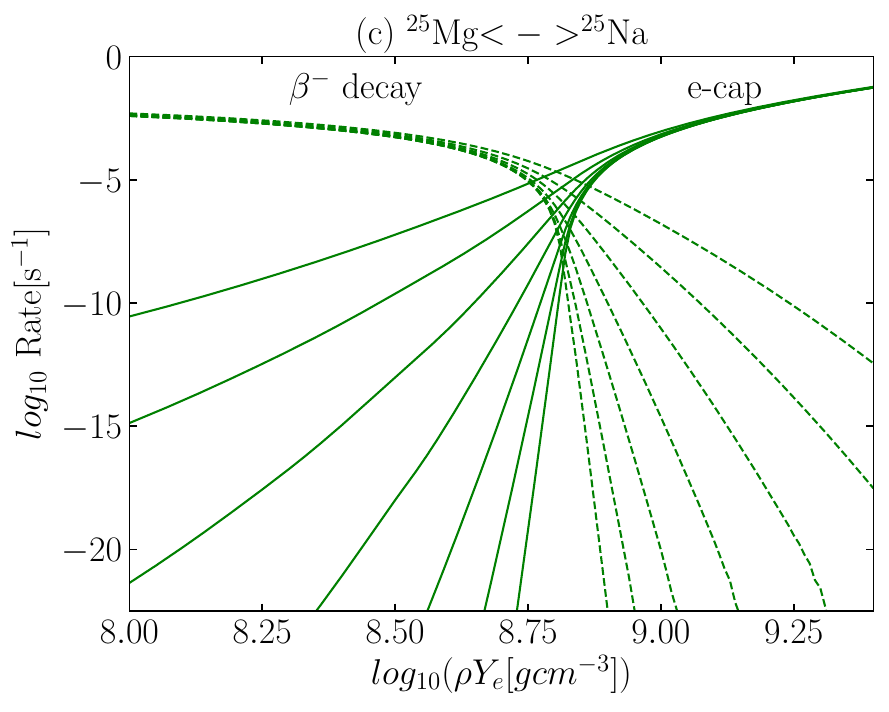}
	\caption{Electron capture and $\beta^-$ decay rates for ($^{25}$Mg, $^{25}$Na) at various temperatures $\log_{10}T=8.0-9.2$ in the interval of 0.2 with Coulomb effects using experimental data along with (a) CEFT, (b) DJ16A, and  (c) USDB effective interaction.\label{rate_25}}
\end{figure}

We now calculate the weak interaction rates for the pairs. 
The weak rates are obtained using CEFT, DJ16A, and USDB interactions, and a comprehensive comparison of the results is carried out. In addition to the purely theoretical shell-model calculations, we also incorporated experimental excitation energies and experimental $B$(GT) data, where available, into the shell-model calculations.

First, the electron capture and beta decay rates for the ($^{23}$Na, $^{23}$Ne) pair are evaluated at various densities and temperatures with the three interactions (CEFT, DJ16A, and USDB). The parent states up to $\approx$2 MeV excitation energy are included. The 3/2$^+_{\rm g.s.}$, 5/2$^+$ (0.440 MeV), 7/2$^+$ (2.076 MeV) and 1/2$^+$ (2.391 MeV) states in $^{23}$Na are included for the electron capture process, and the 5/2$^+_{\rm g.s.}$, 1/2$^+$ (1.017 MeV) states in $^{23}$Ne are included for the $\beta^-$ decay. Figures \ref{rate_23} and \ref{rate_23_expt} correspond to ($^{23}$Na, $^{23}$Ne) pair of nuclei, representing the cases without and with the inclusion of the available experimental data, respectively. Both of the plots include screening effects. In these plots, the density is varied from $\log_{10} (\rho Y_e[{\rm g cm}^{-3}])=8.0$ to 9.4, and the plots are generated at different temperatures, $\log_{10} (T[{\rm K}])$ =  8.0-9.2 in steps of 0.2.

As the density increases, the chemical potential also increases, leading to an enhancement in the electron capture rates. On the other hand, the $\beta^-$-decay rates decrease due to the blocking of the electrons. Assignment of the Urca density is possible for the three interactions. The Urca density refers to the density at which the electron capture and beta-decay rates coincide almost independently of the temperature and the rates are large enough \cite{Toki2013}. When the density range with high rate values becomes wide, cooling does not occur. Such a situation usually takes place when the transitions between the ground states are forbidden.

Figures \ref{rate_23}(a)-\ref{rate_23}(c) show electron capture and $\beta^-$-decay rates calculated using CEFT, DJ16A and USDB effective interactions, respectively. Although the calculated weak rates for DJ16A are a bit larger than those of the others because of larger GT strength in the low excitation energy region, the densities where electron capture and $\beta$-decay rates coincide are found to be almost independent of the temperature for all the interactions. This temperature insensitivity allows for the clear identification of the Urca densities for each interaction. The Urca densities can be assigned to be be $\log_{10}(\rho Y_e)$ = 8.96($\pm$0.01), 8.97($\pm$0.01), and 8.96($\pm$0.01) for CEFT, DJ16A and USDB, respectively.

Furthermore, in Figs. \ref{rate_23_expt}(a)-\ref{rate_23_expt}(c), the recent available experimental excitation energies and $B$(GT) values are incorporated alongside the shell-model excitation energies and GT
strengths calculated using the CEFT, DJ16A, and USDB effective interactions, respectively. From these plots, it is found that with the inclusion of available experimental data in the shell-model calculations the weak rates become almost independent of the interaction employed. Additionally, one can find the Urca density, common to the three interactions, which is assigned as $\log_{10}(\rho Y_e)$ = 8.96$(\pm$0.01.)

This feature is more clearly shown in Fig. \ref{rate_23na}, where the e-capture and $\beta^-$ decay rates, along with their product are calculated at $\log_{10} T$ = 8.7 and 8.0 with the three effective interactions. In this figure, the corresponding rates are plotted without and with the inclusion of screening effects. This figure highlights how screening modifies the rates, particularly near the Urca density.

In Fig. \ref{rate_23na}(a), 
the weak rates at $\log_{10} T =8.7$ obtained by pure theoretical shell-model calculations without the inclusion of available experimental data are plotted for the three effective interactions. Although slight differences between the interactions are noticed, they disappear when available experimental data for excitation energies and GT strengths are included, as shown in Fig. \ref{rate_23na}(b). Figures \ref{rate_23na}(b) and \ref{rate_23na}(c) show the weak rates at $\log_{10} T$ = 8.7 and 8.0, respectively, using the latest available experimental data for excitation energies and GT strengths. The rates calculated using the three effective interactions are almost the same as was observed in Fig.  \ref{rate_23_expt}. The Urca density assigned at $\log_{10} \rho Y_e= 8.96$ is shifted to a lower density at $\log_{10} \rho Y_e= 8.92$ in the case without the screening effects.

We thus found that the three interactions give almost equivalent rates with the same value for the Urca density when available experimental data are taken into account in the shell model calculations. This is mainly because the most important
contributions come from the g.s. to g.s. transitions, for which the experimental data are available. 

The weak interaction rates for the ($^{25}$Mg, $^{25}$Na) pair evaluated for the case with available experimental data are shown in Fig. \ref{rate_25}. The 5/2$^+_{\rm g.s.}$, 1/2$^+$ (0.585 MeV), 3/2$^+$ (0.975 MeV), 7/2$^+$ (1.612 MeV) and 5/2$^+_1$ (1.965 MeV) states in $^{25}$Mg and 5/2$^+_{\rm g.s.}$, 3/2$^+$ (0.0895  MeV), 1/2$^+$ (1.069 MeV) states in $^{25}$Na are included for the parent states. The rates are plotted for the density range $\log_{10} (\rho Y_e) = 8.0-9.2$
and at the different temperatures $\log_{10}(T) =8.0-9.2$ varied in the interval of $\log_{10}(T) =0.2$. It is evident from these plots, similar to the case of the ($^{23}$Na, $^{23}$Ne) pair, when we take into account the available experimental data in the shell-model calculations, there is not much difference in the calculated rates among the shell-model interactions used. From these plots, the Urca density is found to be $\log_{10} \rho Y_e=  8.82 (\pm 0.01)$.

We now discuss the ($^{25}$Na, $^{25}$Ne) pair, which plays an important role in cooling the O-Ne core in the late stages of the evolution of the core \cite{schwab2017}. For the parent states, the 5/2$^+_{\rm g.s.}$, 3/2$^+$ (0.0895 MeV), and 1/2$^+$ (1.069 MeV) states in $^{25}$Na and 1/2$^+_{\rm g.s.}$,  3/2$^+$ and 5/2$^+$ states in $^{25}$Ne are included. Although the g.s. to g.s. transition is forbidden, the transition from the $^{25}$Na (3/2$^{+}_1$) state to the $^{25}$Ne (1/2$^{+}$, g.s.) state is an allowed GT transition. The excitation energy of the $^{25}$Na(3/2$^{+}_1$) state is quite low and the $B$(GT) value of the transition is rather high: $B$(GT) = 0.074, which makes this transition a dominant contributor to the weak rates. This can lead to assigning the Urca density for the three effective interactions.

\begin{figure}[h]\label{rate_25naa_expt}
 \includegraphics[width=77mm]{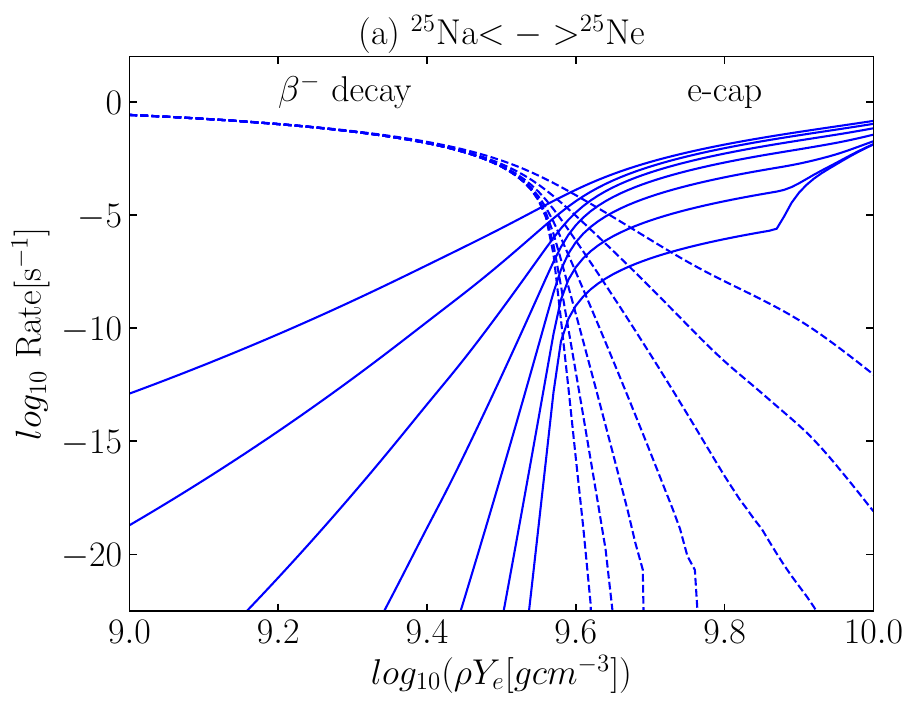}
 \includegraphics[width=77mm]{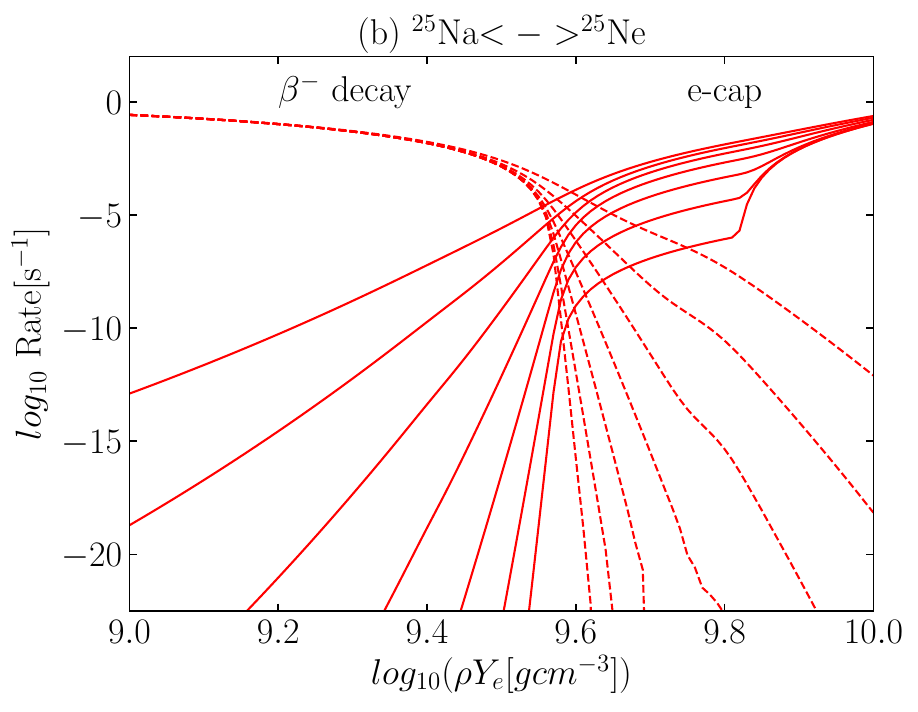}
 \includegraphics[width=77mm]{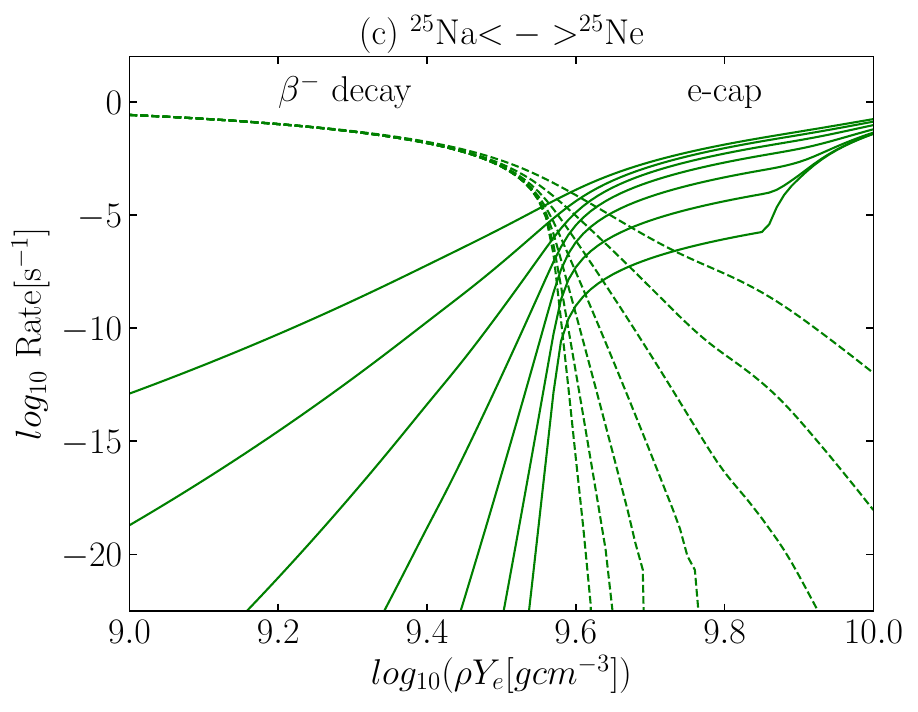}
	\caption{Electron capture and $\beta^-$ decay rates for ($^{25}$Na, $^{25}$Ne) at various temperatures $\log_{10}T=8.0-9.2$ in the interval of 0.2 with Coulomb effects using experimental data along with (a) CEFT, (b) DJ16A, and  (c) USDB effective interactions.\label{rate_25naa_expt}}
\end{figure}

Figure \ref{rate_25naa_expt} shows the weak interaction rates for this nuclear pair at different temperatures $\log_{10}T$ = 8.0-9.2 in steps of 0.2 and density range $\log_{10}(\rho Y_e)=9.0-10.0$ 
with the inclusion of recent available experimental data. 

It is observed from Fig. \ref{rate_25naa_expt} that the weak rates become nearly independent of the effective interactions employed once the experimental data are incorporated. Sudden jumps in the electron capture rates at $\log_{10}(\rho Y_E) \approx$ 9.9 at lower temperatures correspond to the trigger of the transition $^{25}$Na (5/2$_{\rm g.s.}^{+}$) $\rightarrow$  $^{25}$Ne (3/2$^{+}$). The Urca density can be assigned at $\log_{10}(\rho Y_e)$ = 9.59($\pm$0.01) at high temperatures ($\log_{10}(T) \ge$ 8.4) for all the interactions. This density corresponds to the $^{25}$Na (3/2$^{+}$, 0.0895 MeV) $\leftrightarrow$ $^{25}$Ne (1/2$^{+}$,  g.s.) transition.

\begin{figure}\label{rate_29}
 \includegraphics[width=77mm]{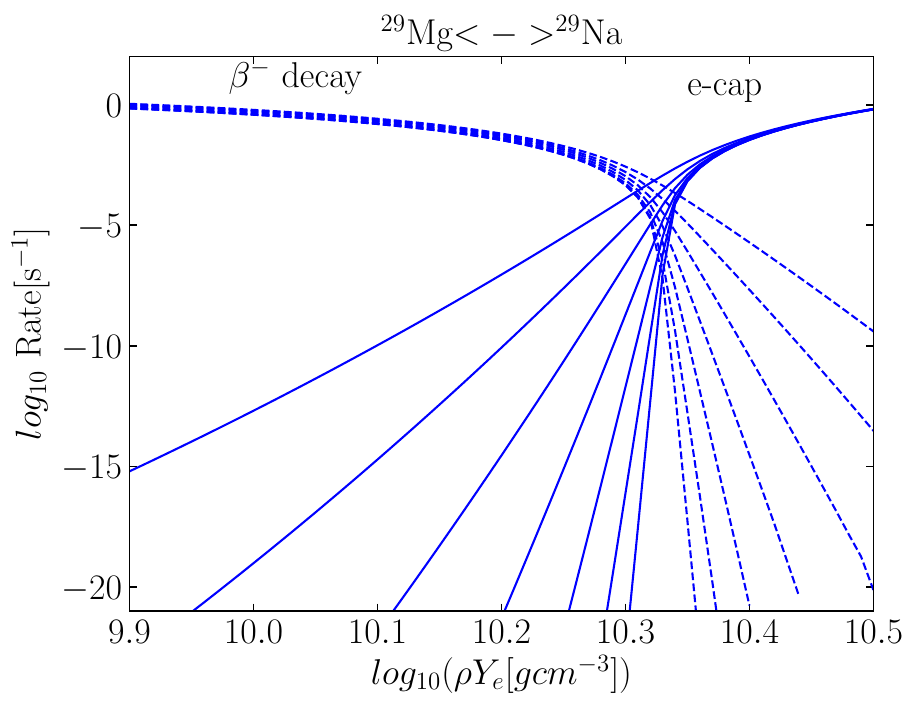}

	\caption{Electron capture and $\beta^-$ decay rates for ($^{29}$Mg, $^{29}$Na) at various temperatures $\log_{10}T=8.0-9.2$ in the interval of 0.2 with Coulomb effects using CEFT effective interaction along with available experimental data.\label{rate_29}}
\end{figure}

\begin{figure*}\label{bgt_31al}
\centering
 \includegraphics[width=82mm]{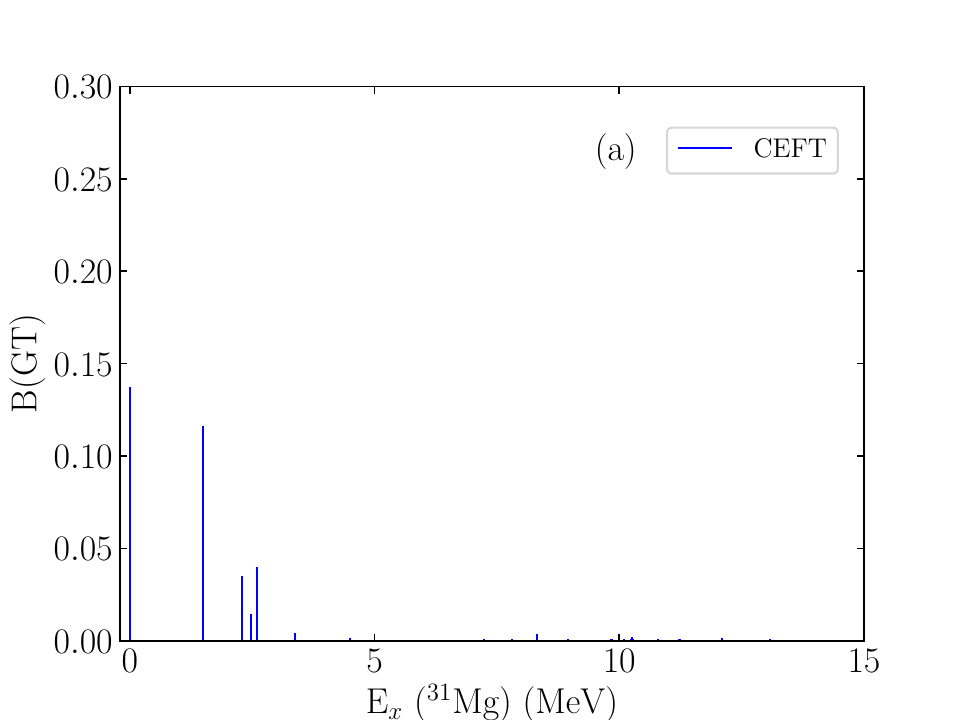}
 \includegraphics[width=82mm]{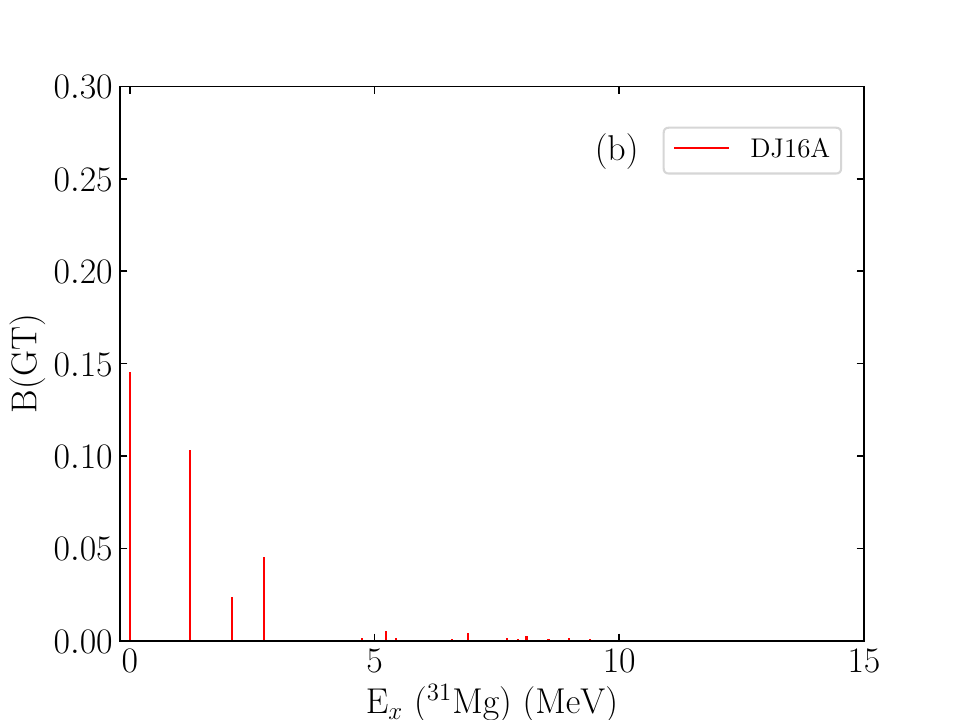}
 \includegraphics[width=82mm]{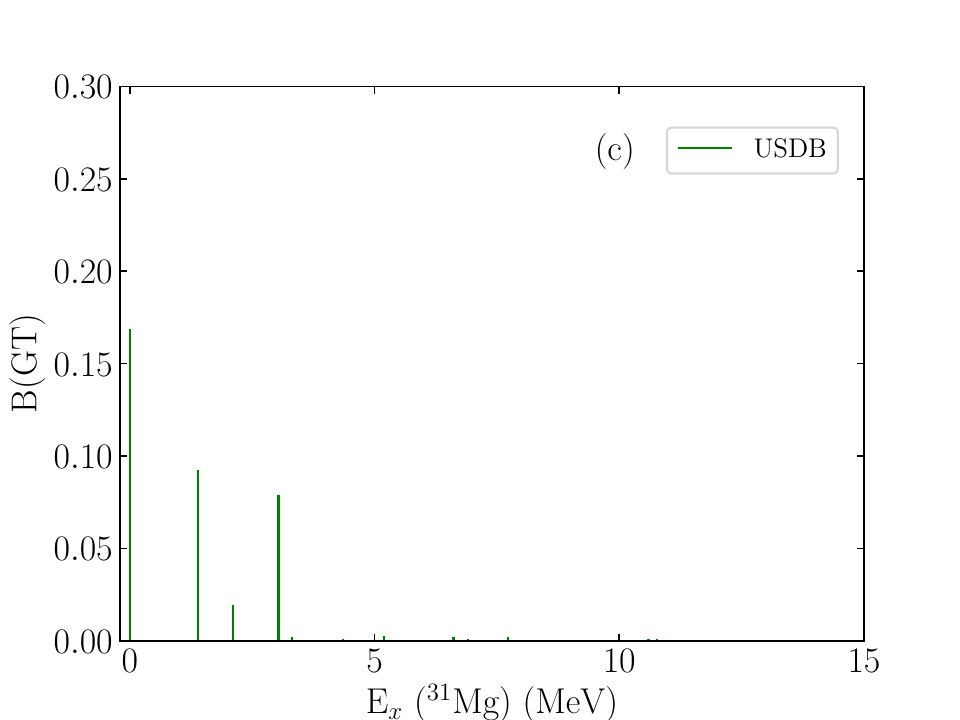}
 \includegraphics[width=82mm]{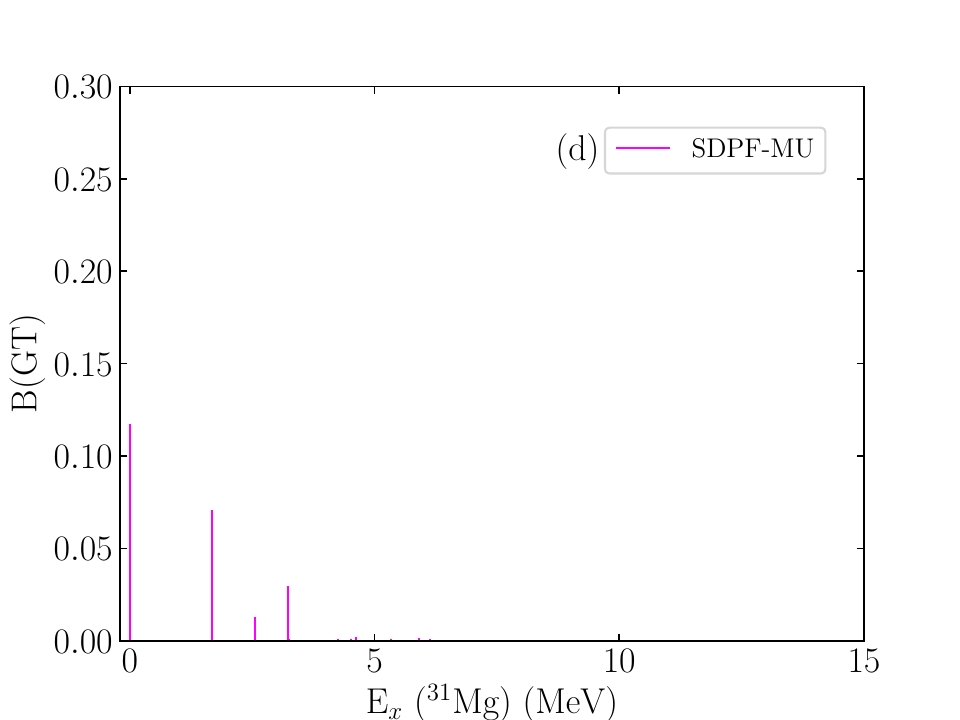}
 \includegraphics[width=82mm]{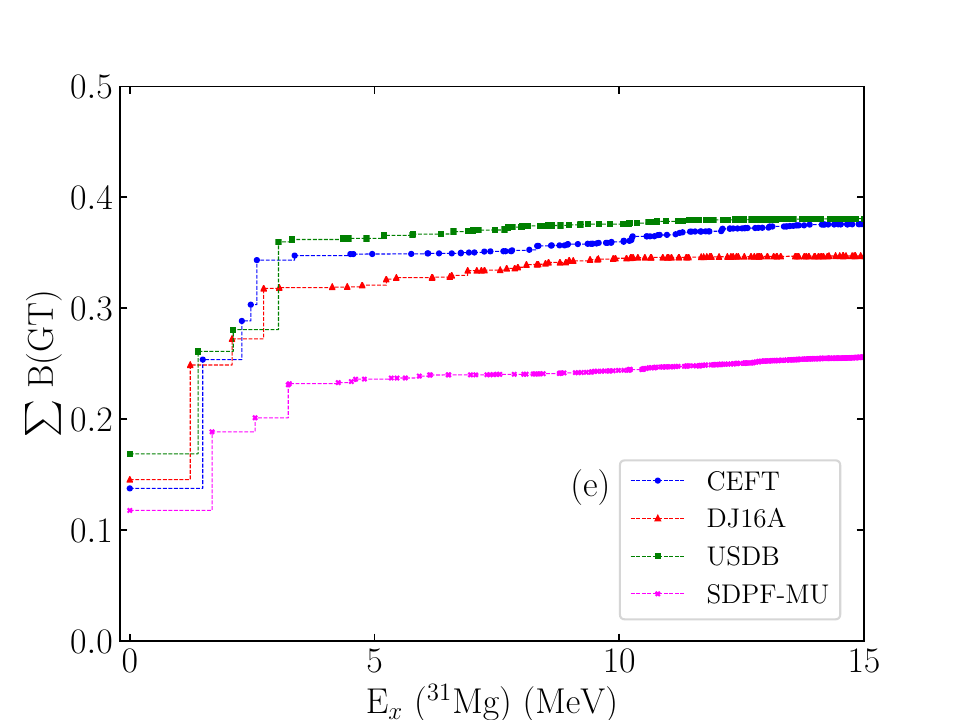}
\caption{$B$(GT) and $\sum B$(GT) values from $5/2^+_{\rm g.s.}$ of $^{31}$Al to various excited states of $^{31}$Mg calculated using CEFT, DJ16A, USDB, and SDPF-MU   effective interactions.\label{bgt_31al}}
\end{figure*}

\begin{figure}[hbt]\label{rate_31}
 \includegraphics[width=77mm]{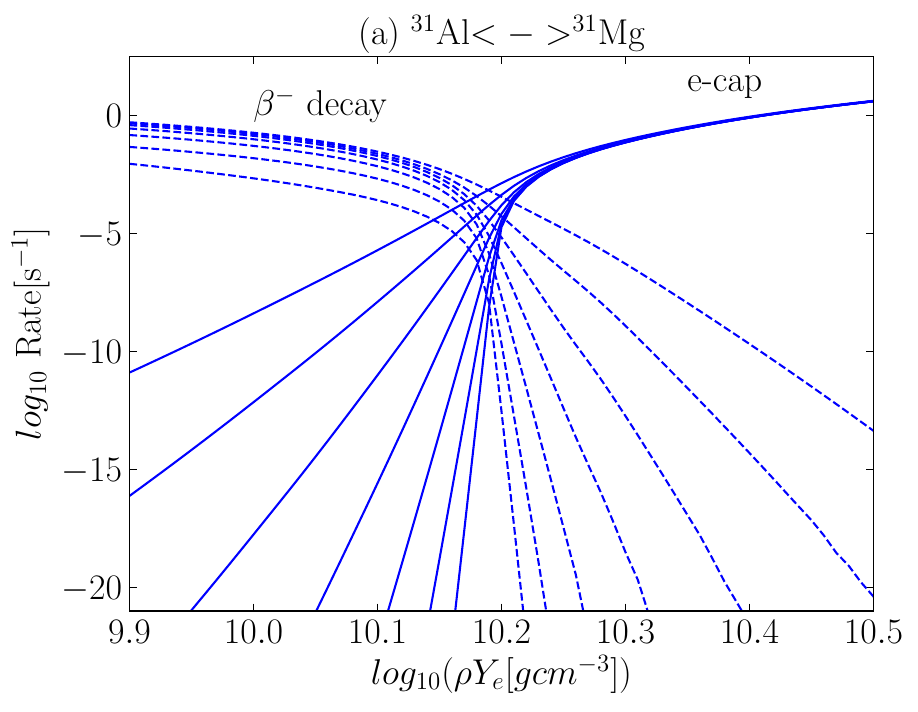}
\includegraphics[width=77mm]{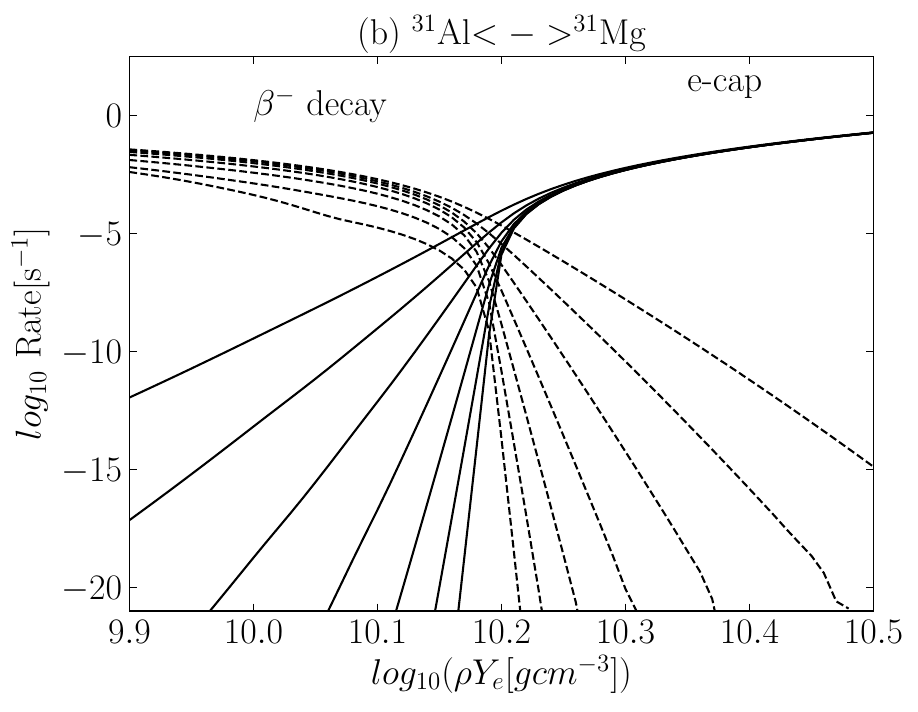}
	\caption{Electron capture and $\beta^-$ decay rates for ($^{31}$Al, $^{31}$Mg) at various temperatures $\log_{10}T=8.0-9.2$ in the interval of 0.2 with Coulomb effects using (a) CEFT and (b) EEdf1 effective interactions along with available experimental data.\label{rate_31}}
\end{figure}
\begin{figure}[hbt]\label{rate_31al}
 \includegraphics[width=78mm]{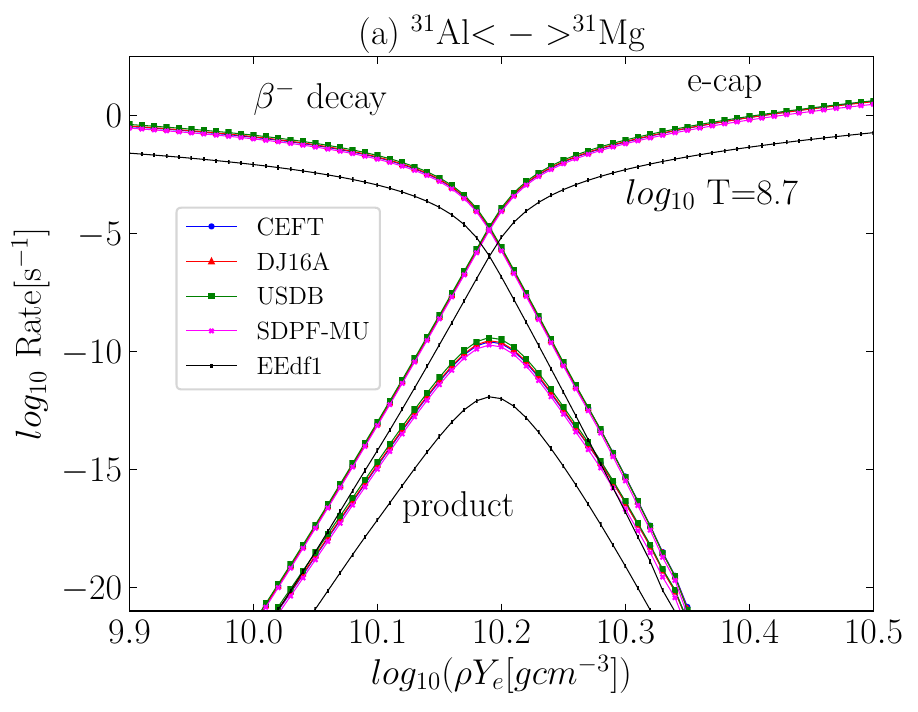}
 \includegraphics[width=78mm]{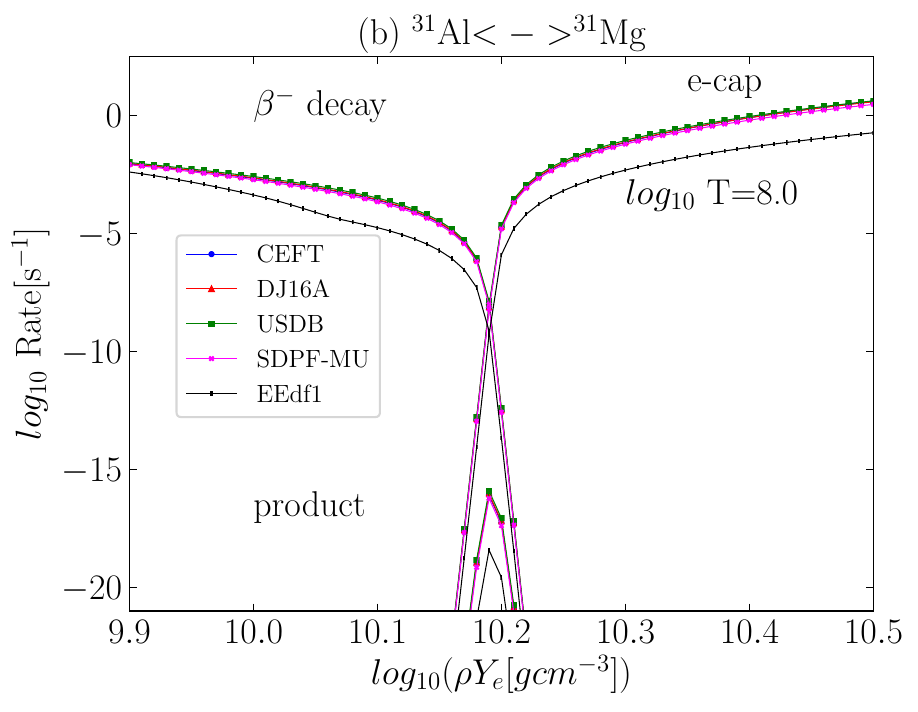}
	\caption{Electron capture and $\beta^-$ decay rates for ($^{31}$Al, $^{31}$Mg) at (a) $\log_{10}T=8.7$ and (b) $\log_{10}T=8.0$ with Coulomb effects using CEFT, DJ16A, USDB, SDPF-MU and EEdf1 effective interactions. Available experimental data are included.
    }
    \label{rate_31al}
\end{figure}

\begin{figure}\label{rate_31pp}
 \includegraphics[width=77mm]{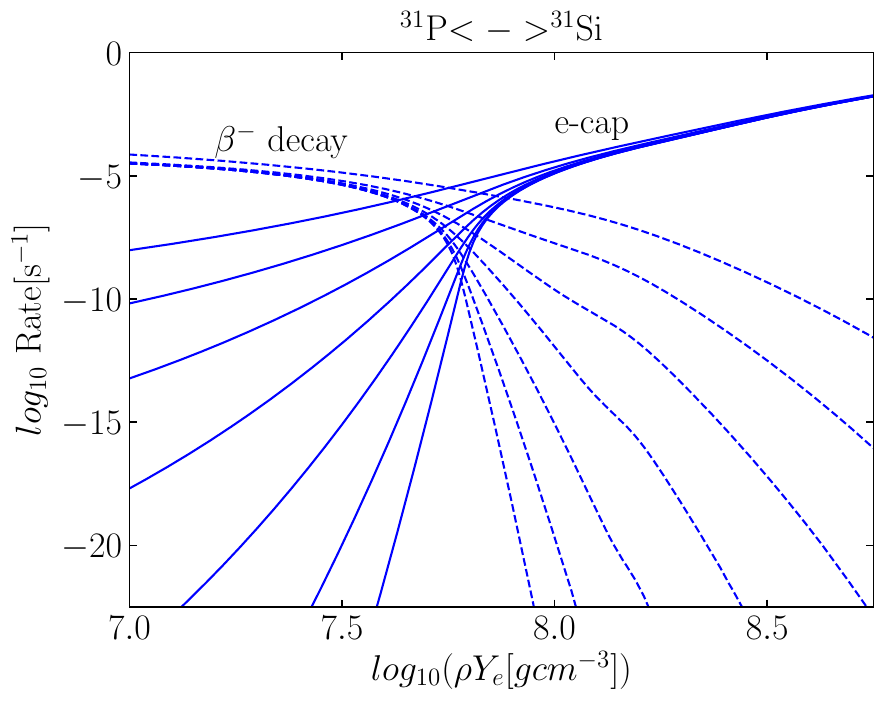}
 \caption{Electron capture and $\beta^-$ decay rates for ($^{31}$P, $^{31}$Si) at various temperatures $\log_{10}T=8.0-9.2$ in the interval of 0.2 with Coulomb effects using CEFT effective interaction along with available experimental data.\label{rate_31pp}}
\end{figure}

\begin{figure}[hbt]\label{rate_33alrate}
\includegraphics[width=82mm]{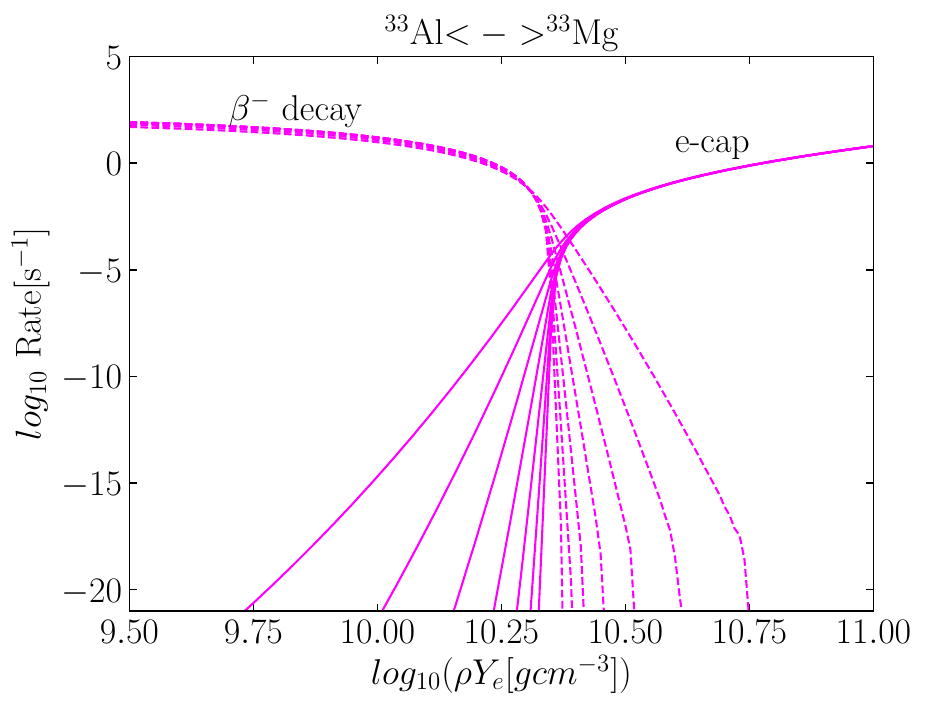}
\caption{Electron capture and $\beta^-$ decay rates for ($^{33}$Al, $^{33}$Mg) at various temperatures $\log_{10}T=8.0-9.2$ in the interval of 0.2 with Coulomb effects using SDPF-MU effective interaction along with available experimental data.\label{rate_33alrate}}
\end{figure}

\subsection{Gamow-Teller strengths and weak rates for Urca pairs with $A$ = 29, 31 and 33}

We have also calculated weak interaction rates for nuclei with mass numbers $A=29$, $31$ and $33$, which include ($^{29}$Mg, $^{29}$Na), ($^{31}$Al, $^{31}$Mg), ($^{31}$P, $^{31}$Si), and ($^{33}$Al, $^{33}$Mg) nuclear pairs. The ($^{31}$Al, $^{31}$Mg) pair was pointed out to be important in cooling neutron star crusts \cite{schatz2014,deibel2016,ong2020}. 
The ($^{29}$Mg, $^{29}$Na) and ($^{33}$Al, $^{33}$Mg) pairs, which have $Q$ value similar to the ($^{31}$Al, $^{31}$Mg) pair, are other candidates for Urca cooling in the neutron star crusts \cite{schatz2014}. The $\beta$-decay $Q$ value for the($^{31}$P, $^{31}$Si) pair is as small as 1.49 MeV, and the pair is expected to play some important roles in the cooling and neutralization of the core in later O-burning and Si-burning stages in massive stars.

For the calculation of the weak rates of the ($^{29}$Mg, $^{29}$Na) pair, the 3/2$^+_{\rm g.s.}$, 1/2$^+$, and 5/2$^+$ states of $^{29}$Mg and 3/2$^+_{\rm g.s.}$ and 5/2$^+$  (0.072 MeV) states of $^{29}$Na are included as the parent states. The experimental GT strength between the ground states is available; $B$(GT) =0.038 \cite{Mueller1984}. 
Figure \ref{rate_29} shows the electron capture and $\beta^-$-decay rates at different temperatures ($\log_{10}T$ =  8.0-9.2 in steps of 0.2) and density range $\log_{10}(\rho Y_e) = 9.9-10.5$ using the CEFT interaction with the inclusion of the available experimental data. The Urca density in this case can be assigned at  $\log_{10}(\rho Y_e)$ = $10.33(\pm0.01)$. The CEFT and USDB effective interactions show similar rate trends, whereas the DJ16A effective interaction shows a slight difference in the rates from the rest of the two. Nevertheless, the Urca density comes out to be $\log_{10}(\rho Y_e)$ = 10.33($\pm$0.01) (10.28$\pm$0.01) with (without) the screening effects common to the three effective interactions. 

Another effective interaction, i.e., SDPF-MU, is included for the calculation of GT strengths to extend the $sd$-shell model space towards $sd$-$pf$ shell model space for the nuclei with $A=31$ and 33. There are low-lying negative parity states in $^{31}$Mg and $^{33}$Mg.
In Figs. \ref{bgt_31al}(a)-\ref{bgt_31al}(d), the $B$(GT) values from $5/2^+_{\rm g.s.}$ of $^{31}$Al to all possible GT transitions in $^{31}$Mg up to 20 MeV excitation energy, calculated by CEFT, DJ16A, USDB and SDPF-MU effective interactions are plotted, respectively. In these plots, a prominent peak can be seen for the transition to the $3/2^+_{1}$ of $^{31}$Mg  which is maximum, $B$(GT) =0.17, for the USDB and minimum, $B$(GT) =0.12, for the SDPF-MU, within $\approx40\%$ difference. 
Further, Fig. \ref{bgt_31al}(e) shows the cumulative sum of GT strengths for all these effective interactions, which almost saturates after excitation energy of 10 MeV.

Next, for the ($^{31}$Al, $^{31}$Mg) pair, the 5/2$^+_{\rm g.s.}$, 1/2$^+$ (0.947 MeV) and 3/2$^+$ (1.613 MeV) states of $^{31}$Al and 1/2$^+_{\rm g.s.}$, and 3/2$^+$ (0.050 MeV)  states of $^{31}$Mg are included as the parent states for the calculation of the weak rates. In this case, the SDPF-MU and EEdf1 \cite{tsunoda2017,tsunoda2020} effective interactions are also used along with CEFT, DJ16A and USDB effective interactions. These effective interactions are used to include higher configurations in the wave function calculations. Up to two-particle two-hole (2p-2h) and six-particle six-hole (6p-6h) excitations from the $sd$ shell to the $pf$ shell are included in the SDPF-MU and EEdf1 effective interactions, respectively. No quenching factor in the coupling constant, i.e., $q_A=1.0$ is used for the GT strengths in the EEdf1 effective interaction due to the extended model space. All effective interactions  except for the EEdf1 predict $3/2^+$ as the g.s. in $^{31}$Mg. The EEdf1 predicts 1/2$^+$ as the g.s. and 3/2$^{+}$ as the first excited state with the excitation energy of 0.052 MeV. 
Negative parity states are calculated to be at 137 keV (3/2$^{-}$) and 571 keV (7/2$^{-}$) for the EEdf1 close to the experimental data at 221 keV and 461 keV, respectively. The SDPF-MU, on the other hand, does not predict low-lying negative parity states. 
Here GT transitions between positive parity states are considered. Figures \ref{rate_31}(a) and \ref{rate_31}(b) show the weak rates in density range $\log_{10}(\rho Y_e) = 9.9-10.5 $ at different values of temperature $\log_{10}T =  8.0-9.2$ in steps of 0.2 evaluated using CEFT and EEdf1 effective interactions with the inclusion of the available experimental data, respectively. The Urca density can be assigned as 
$\log_{10}(\rho Y_e)$ = 10.19 ($\pm0.02$) for both of the interactions. Further, Fig. \ref{rate_31al} shows the weak interaction rates at $\log_{10}T$=8.7 and 8.0 with the inclusion of the available experimental data. The most important contributions come from the GT transitions between the 5/2$_{\rm g.s.}^{+}$ state in $^{31}$Al and the 3/2$_{1}^{+}$ (0.050 MeV) state in $^{31}$Mg. Calculated $B$(GT)'s obtained by the CEFT, DJ16A, USDB and SDPF-MU interactions, $B$(GT: $^{31}$Al $\rightarrow$ $^{31}$Mg) = 0.12-0.17, do not differ more than $\approx40\%$. Thus, there is not much variation in the weak interaction rates calculated via all of the effective interactions in both of the plots except for the rates calculated with the EEdf1. The rates with the EEdf1 are suppressed as compared to the other ones because of the low GT strengths, for example, $B$(GT: $^{31}$Al (5/2$_{\rm g.s.}^{+}$) $\rightarrow$ $^{31}$Mg (3/2$^{+}$, 0.05 MeV)) = 0.0094. Nevertheless, the Urca density remains the same for all of the effective interactions. The Urca density with and without the inclusion of the screening effects comes out to be 10.19 and 10.14, respectively. Here, only the weak decay rates with the inclusion of the screening effects are shown. The weak rates evaluated by the EEdf1 interaction without the Coulomb effects and without available experimental data are shown in Refs. \cite{ssfr2024,suzuki2022}.

 Next, we discuss the ($^{31}$P, $^{31}$Si) pair. The experimental $B$(GT) value is only available for the transition from the $1/2^+_{\rm g.s.}$ state of $^{31}$P to the $3/2^+_{\rm g.s.}$ state of $^{31}$Si and it matches well with the shell-model calculations. The spread of these GT strengths is seen up to approximately 10 MeV excitation energy of $^{31}$Si for all of the effective interactions, and their strength starts to diminish after 10 MeV. The cumulative sum of the GT strengths calculated via all of the above-mentioned effective interactions shows a saturation after 10 MeV excitation energy of $^{31}$Si.

For the calculation of the weak rates of the ($^{31}$P, $^{31}$Si) pair, the 1/2$^+_{\rm g.s.}$, and 3/2$^+$ (1.266 MeV) states of $^{31}$P and 3/2$^+_{\rm g.s.}$, 1/2$^+$ (0.752 MeV) and 5/2$^+$ (1.695 MeV) states of $^{31}$Si are included as the parent states.
The CEFT, DJ16A, USDB, and SDPF-MU effective interactions are used for these weak rate calculations  along with the available experimental data. Similar to Fig. \ref{rate_29}, the electron capture and $\beta^-$-decay rates are evaluated at different temperatures using the CEFT effective interaction in the density range $\log_{10}(\rho Y_e)$ = 7.0-8.75 as shown in Fig. \ref{rate_31pp}. 
The Urca density can be assigned as 
$\log_{10}(\rho Y_e)$ = 7.77 ($\pm$ 0.02). This value of the Urca density proves to be common to all the effective interactions when the experimental data are included.

Moving forward, for the calculation of the weak rates for the ($^{33}$Al, $^{33}$Mg) pair, the 5/2$^+_{\rm g.s.}$ state of $^{33}$Al, the 3/2$^-_{\rm g.s.}$, 7/2$^-$ (0.154 MeV), 3/2$^+$ (0.484 MeV) and 3/2$^-$ (0.703 MeV) states of $^{33}$Mg are included as the parent states.
Although spin-parities of the excited states in $^{33}$Mg are not confirmed, they are taken as above \cite{Bazin2021}. 
In this case, the SDPF-MU effective interaction is used along with the available experimental data. First, the electron capture and $\beta^-$-decay rates are calculated at different temperatures $\log_{10}T$= 8.0-9.2 in steps of 0.2 in density range $\log_{10} \rho Y_e=9.5-11.0$ in fine grid. Here, g.s. to g.s. transition is forbidden thus Urca density can not be clearly assigned. However, we have included the contribution of g.s. to g.s. forbidden transition by using an experimental $\log ft$ value,  $\log ft$ =5.6 for the $\beta$ decay $^{33}$Mg (3/2$^{-}$) $\rightarrow$ $^{33}$Al (5/2$^{+}$), and calculating $B$(GT) value from this $\log ft$ value by assuming it as allowed transition. The rest of the transitions included are GT transitions and because of the lack of the experimental data, the GT values used are calculated via SDPF-MU effective interaction. Thus, we can make an estimate for Urca density from Fig. \ref{rate_33alrate} at $\log_{10}\rho Y_e=10.36(\pm0.02)$.

\subsection{Effects of the Urca pairs with $A$ = 29, 31 and 33 to the cooling of neutron star crusts}

In this subsection, we apply the weak rates of nuclear pairs
obtained here to the cooling of neutron star crusts by the nuclear Urca process. The neutrino luminosity of the Urca process can be expressed as  \cite{deibel2016,tsuruta1970}
\begin{equation}
L_{\nu} \approx L_{34}\times10^{34} ({\rm erg/s}) X(A) T_9^5 (\frac{g_{14}}{2})^{-1} R_{10}^2,
\end{equation}
where $X(A)$ is the mass fraction of the parent nucleus, $T_9$ is the crust temperature in GK, $g_{14}$ is the neutron star surface gravity in units of 10$^{14}$ cm s$^{-2}$, and  $R_{10}$ is the neutron star radius in units of 10 km. Effects of nuclear properties are given by the intrinsic cooling strength L$_{34}$.

The formula for determination of intrinsic cooling strength, i.e., L$_{34}$ by considering thermally populated excited states is given by \cite{wang2021}

\begin{align}
L_{34}(Z,A,T) = \sum_{\epsilon\beta} \, & 0.87 \left( \frac{10^6~\text{s}}{\langle ft \rangle_{\epsilon\beta}} \right)
\left( \frac{56}{A} \right) \nonumber \\
& \times \left( \frac{|Q_{\epsilon\beta}(Z,A)|}{4~\text{MeV}} \right)^5
 \left( \langle F \rangle^+ \right),
\end{align}
where $Z$ and $A$ are the proton and mass number of the EC parent nucleus and $\epsilon$ and $\beta$ denotes the EC and $\beta^-$-decaying nucleus. The $\langle F\rangle^+$ is the Coulomb correction induced by electron-ion interaction given by

\begin{equation}
\langle F\rangle^{+}= \frac{2\pi \alpha Z}{|1-e^{(-2\pi \alpha Z)}|} .
\end{equation}
 $\langle F\rangle^{+}$, which is an approximation to the Fermi function, is larger than unity because of attraction between electron and nuclei.

Now,  $Q_{\epsilon\beta}(Z,A)$ is the $Q$ value of the corresponding transition and it is given as
\begin{equation}
Q_{\epsilon\beta}(Z,A)=Q_{\rm EC}-m_ec^2+E_{\epsilon}-E_{\beta},
\end{equation}
where $Q_{\rm EC}$ is the $Q$ value of the EC parent nucleus and
$E_{\epsilon}$ and $E_{\beta}$ are the excitation energies of the corresponding states of the transitions in the EC and $\beta$-decaying parent nucleus. However, this expression changes when screening effects are considered and $Q_{\epsilon\beta}(Z,A)$ is given by
\begin{equation}
Q_{\epsilon\beta}(Z,A)=Q_{\rm EC}-m_ec^2+E_{\epsilon}-E_{\beta}+\Delta Q_C,
\end{equation}
 where $\Delta Q_C$ is the shift of the $Q$ value given by Eq. \ref{delq}.

Further,  
\begin{equation}
\langle ft\rangle_{\epsilon\beta}=\frac{\tilde{f}t^-_{\epsilon\beta}+\tilde{f}t^+_{\epsilon\beta}}{2}
\end{equation}
where
\begin{equation}
\tilde{f}t^+_{\epsilon\beta}=\frac{G^+(T)}{(2J_{\epsilon}+1)e^{-E_{\epsilon}/kT}}ft^+_{\epsilon\beta},
\end{equation}
and
\begin{equation}
\tilde{f}t^-_{\epsilon\beta}=\frac{G^-(T)}{(2J_{\beta}+1)e^{-E_{\beta}/kT}}ft^-_{\epsilon\beta}.
\end{equation}

Here, $G^{\pm}(T)=\sum_{\epsilon/\beta}(2J_{\epsilon/\beta}+1)e^{-E_{\epsilon/\beta}/kT}$ is the partition function of EC/$\beta^-$ parent nucleus, and $ft^+_{\epsilon\beta}$ ($ft^-_{\epsilon\beta} $) is the $ft$ value for EC ($\beta^{-}$) process.

\begin{figure}[hbt]\label{l34_29mg}
\includegraphics[width=82mm]{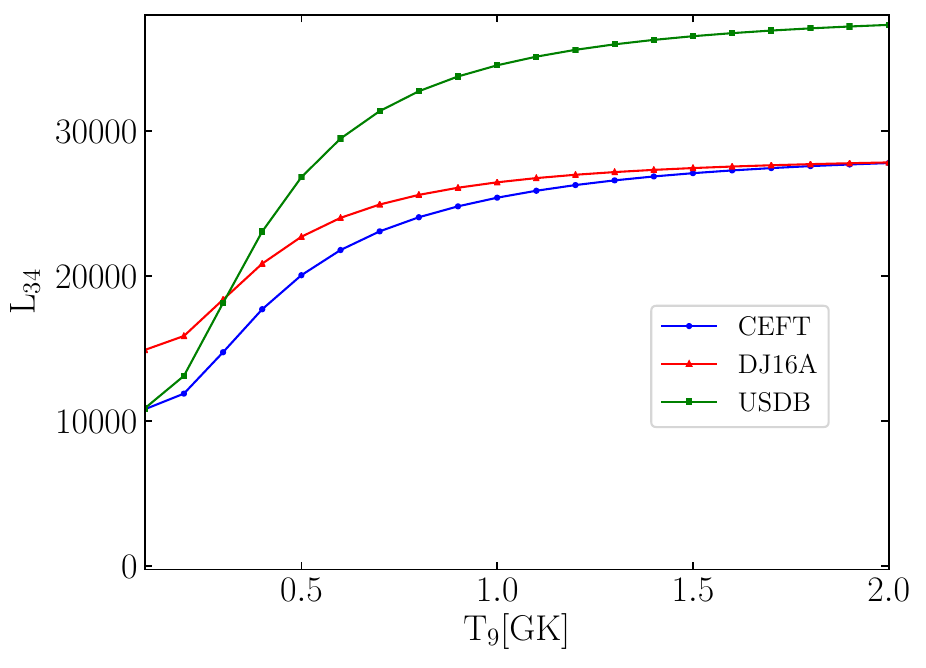}
\caption{Intrinsic cooling strength ($L_{34}$) for ($^{29}$Mg, $^{29}$Na) with Coulomb effects 
using CEFT, DJ16A, and USDB effective interactions along with experimental data. \label{l34_29mg}}
\end{figure}

\begin{figure}[hbt]\label{lum_29mg}
\includegraphics[width=82mm]{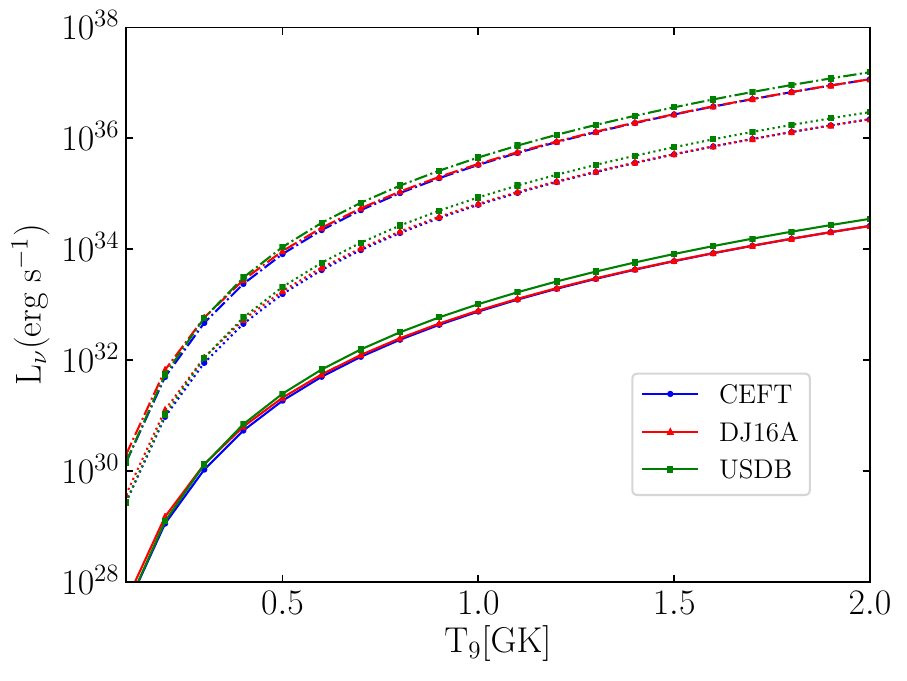}
\caption{Neutrino luminosity ($L_{\nu}$(erg s$^{-1}$)) for ($^{29}$Mg, $^{29}$Na) with Coulomb effects using CEFT, DJ16A, and USDB effective interactions along with experimental data where solid, dotted and dashed-dotted lines corresponds to superbursts, stable hydrogen burning and type I X-ray bursts, respectively. \label{lum_29mg}}
\end{figure}

\begin{figure}[hbt]\label{l34_31al}
\includegraphics[width=81mm]{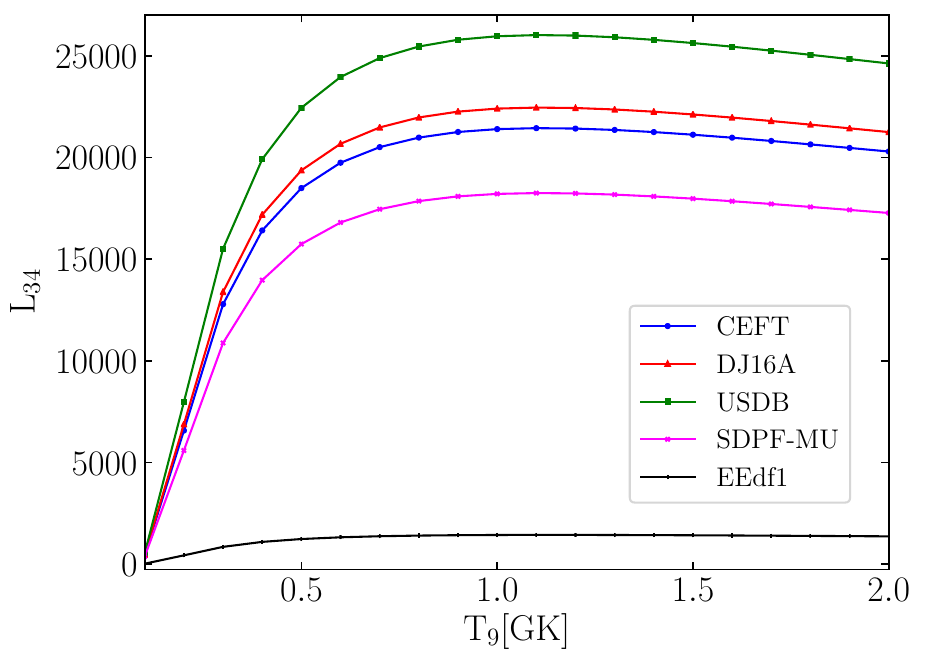}
\caption{Intrinsic cooling strength ($L_{34}$) for ($^{31}$Al, $^{31}$Mg) with Coulomb effects 
using CEFT, DJ16A, USDB, SDPF-MU and EEdf1 effective interactions along with experimental data. \label{l34_31al}}
\end{figure}

\begin{figure}[hbt]\label{lum_31al}
\includegraphics[width=81mm]{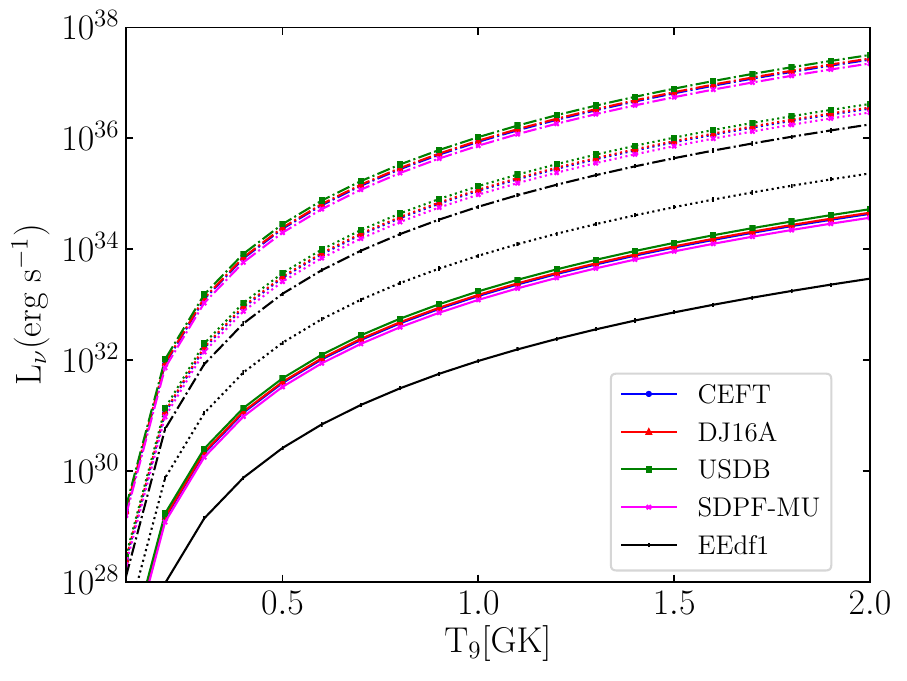}
\caption{
 The same as in Fig. \ref{lum_29mg} for the 
($^{31}$Al, $^{31}$Mg) pair.} 
\label{lum_31al}
\end{figure}

\begin{figure}[hbt]\label{l34_33al}
\includegraphics[width=81mm]{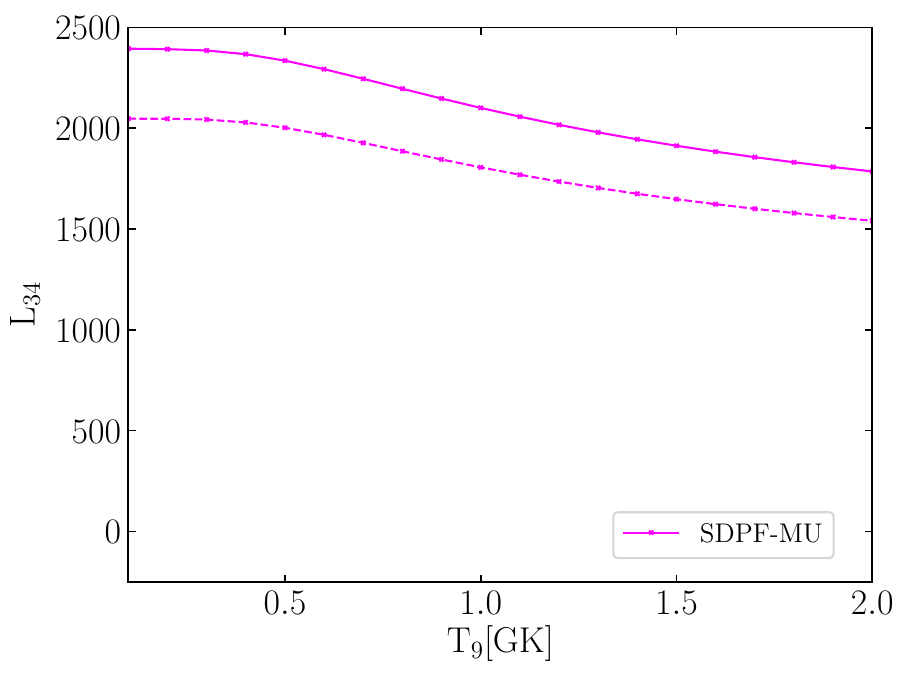}
\caption{Intrinsic cooling strength ($L_{34}$) for ($^{33}$Al, $^{33}$Mg) without (dashed line) and with (solid line) Coulomb effects 
using SDPF-MU effective interaction along with experimental data. \label{l34_33al}}
\end{figure}

\begin{figure}[hbt]\label{lum_33al}
\includegraphics[width=81mm]{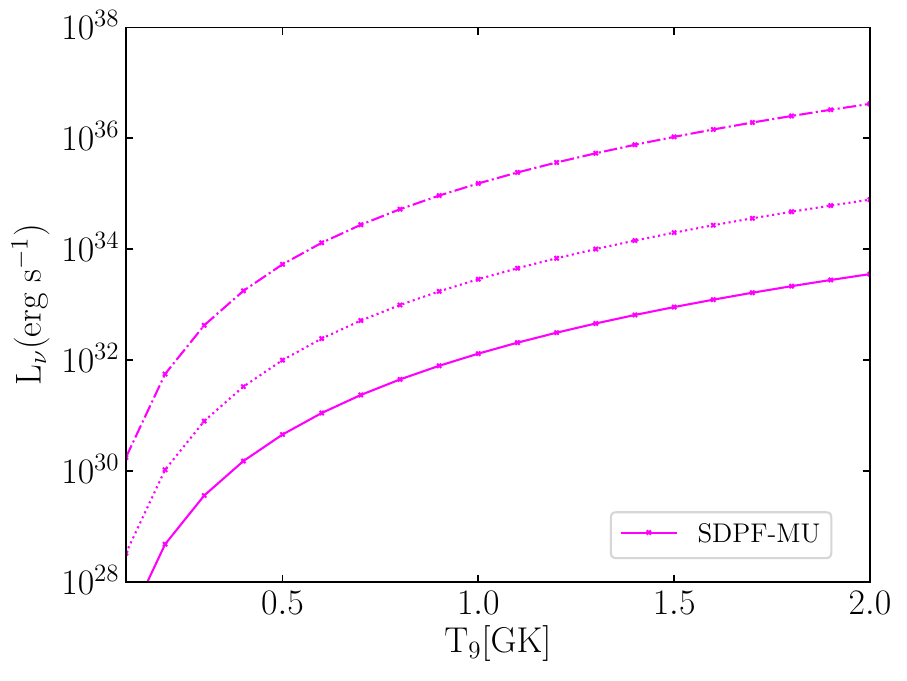}
\caption{Neutrino luminosity ($L_{\nu}$(erg s$^{-1}$)) for ($^{33}$Al, $^{33}$Mg) without (dashed line) and with (solid line) Coulomb effects 
using SDPF-MU effective interaction along with experimental data where solid, dotted and dashed-dotted lines corresponds to superbursts, stable hydrogen burning and type I X-ray bursts, respectively. \label{lum_33al}}
\end{figure}

Using the aforementioned formalism, the neutrino luminosity and intrinsic cooling strength are calculated for the nuclear Urca pairs, ($^{29}$Mg, $^{29}$Na), ($^{31}$Al, $^{31}$Mg) and ($^{33}$Al, $^{33}$Mg). These neutrino luminosities and intrinsic cooling strengths are calculated by incorporating the screening effect
in the corresponding $Q$ value.  Three different mass fraction values, i.e.,  $X(A)$ = $X_{\rm SB}$, $X_{\rm S}$ and $X_{\rm XRB}$ are adopted from Ref. \cite{meisel2017} corresponding to superbursts, stable hydrogen burning, and type I X-ray bursts, respectively. Here, we adopted $g_{14}=1.85$ and $R_{10}$= 1.2.

First, the intrinsic cooling strength and neutrino luminosity are calculated for the ($^{29}$Mg, $^{29}$Na) Urca pair, which was
predicted to give the highest luminosity among the Urca pairs
responsible for the cooling of neutron star crust \cite{schatz2014}. The input data, such as $\log ft$ values, spin parity of states, and the corresponding excitation energies, are calculated from CEFT, DJ16A, and USDB effective interactions along with the experimental data wherever available. The variation of $L_{34}$ with wide range of temperature $T=0-2.0$ GK in fine grids is shown in Fig. \ref{l34_29mg}. Here, the screening effects in $Q$ value at various temperatures are calculated at the Urca density $\log_{10} \rho Y_e=10.33$,  which gives $\Delta Q_C$ $\approx$-0.40 MeV. The $L_{34}$ remains almost constant up to $T_9=0.1$ GK because $^{29}$Mg and $^{29}$Na nuclei remains in their ground states up to $T=0.1$ GK. Since the $1/2^+$ state is predicted at very low excitation energy, almost degenerate with the 3/2$^+_{\rm g.s.}$, for the DJ16A interaction, the intrinsic cooling strength, i.e., $L_{34}$ is enhanced in this case. 
After 0.1 GK temperature, the $5/2^+_1$ state at energy level 72 keV in $^{29}$Na is thermally populated, and its occupation probability is increased. The transition between the first excited state of $^{29}$Na and g.s. of $^{29}$Mg is allowed, and the $L_{34}$ starts to increase after 0.1 GK temperature. Out of the three considered effective interactions, the USDB interaction has the largest matrix element for this transition. Thus, the $L_{34}$ is enhanced for USDB interaction
as compared to the other two interactions. Further, the neutrino luminosity ($L_{\nu}$(erg s$^{-1}$)) is calculated for this Urca pair by using mass fraction values $X_{\rm SB} = 1.9\times 10^{-6}$, $X_{\rm S} = 1.6\times 10^{-4}$, and $X_{\rm XRB} = 8.4\times 10^{-4}$  \cite{meisel2017}. The resulting values of $L_{\nu}$(erg s$^{-1}$) values are plotted in Fig. \ref{lum_29mg}, where the solid, dotted, and dash-dotted curves represent superburst, stable hydrogen burning, and type I X-ray burst conditions, respectively. Because of the highest mass fraction associated with type I X-ray bursts, the neutrino luminosity is maximum in this case. Additionally, the neutrino luminosity is enhanced for the USDB
interaction compared to others because of the large $L_{34}$ value for USDB interaction.

Next, we discuss the ($^{31}$Al,$^{31}$Mg) Urca pair, which was shown to give high cooling luminosity in the neutron star crusts \cite{schatz2014,ong2020}. The intrinsic cooling strength and neutrino luminosity are calculated for this nuclear pair using CEFT,
DJ16A, USDB, SDPF-MU, and EEdf1 effective interactions. The corresponding results for intrinsic cooling strength using different effective interactions are compared with each other as shown in Fig. \ref{l34_31al}. The available experimental data are used for the spin parities of the states with corresponding energies and $\log ft$ values of the transitions wherever available. Otherwise, these data are supplemented by the shell-model calculations via different effective interactions as discussed above. The experimental $\log ft^-$ values for $^{31}$Mg($1/2^+_{\rm g.s.}$) $\rightarrow$ $^{31}$Al($1/2^+_1$, $3/2^+_1$) are 6.02 and 5.59, respectively. The $\log ft^-$ values for $^{31}$Mg($3/2^+_1$) $\rightarrow$ $^{31}$Al($5/2^+_{\rm g.s.}$, $1/2^+_1$) are calculated with the help of shell-model calculations. The value of $\Delta Q_C$ at the Urca density $\log_{10} (\rho Y_e) =10.19$ is around -0.38 MeV. In the ($^{31}$Al,$^{31}$Mg) pair, the g.s. to g.s. transition is forbidden, which makes $L_{34}$ negligible at the starting temperature. With increasing temperature, the first excited state $3/2^+_1$ at 50 keV in $^{31}$Mg gets thermally populated and has considerable occupancy. Now, the allowed transition $^{31}$Mg($3/2^+_1$) $\rightarrow$ $^{31}$Al($5/2^+_{\rm g.s.}$) starts contributing towards the intrinsic cooling strength, which leads to an increase in the $L_{34}$ after 0.1 GK. The GT strength calculated from EEdf1 interaction for $^{31}$Mg($3/2^+_1$) $\rightarrow$ $^{31}$Al($5/2^+_{\rm g.s.}$) is low as compared to those by other effective interactions. Thus, $L_{34}$ is suppressed for EEdf1 interaction as compared to other effective interactions. After $T \approx $ 1.5 GK, the $3/2^-_1$ at 221 keV in $^{31}$Mg gets considerable occupancy, but the transitions between this state and the lower excited states in $^{31}$Al are all forbidden, which leads to a slight decrease in the $L_{34}$ after temperature 1.5 GK. Furthermore, the neutrino luminosity calculated using the effective interactions mentioned above is plotted in Fig. \ref{lum_31al} where the mass fraction values used are $X_{\rm SB} = 4.3\times 10^{-6}$ and $X_{\rm S} = 3.4\times 10^{-4}$, and $X_{\rm XRB} = 2.6\times 10^{-3}$ \cite{meisel2017}.

Similarly, the intrinsic cooling strength for ($^{33}$Al,$^{33}$Mg) Urca pair is calculated by employing SDPF-MU effective interaction. The corresponding results are plotted in Fig. \ref{l34_33al}. In this plot, the intrinsic cooling strength ($L_{34}$) is shown without (dashed line) and with (solid line) the inclusion of the screening effect in the $Q$ value. $\Delta Q_C \approx$ -0.43 MeV at the Urca density $\log_{10}\rho Y_e=10.36$. 
The inclusion of the screening effect enhances the $L_{34}$ by approximately 15$\%$. 
The g.s. to g.s. transition, i.e, $^{33}$Mg($3/2^-_{\rm g.s.}$) $\rightarrow$ $^{33}$Al($5/2^+_{\rm g.s.}$) is first forbidden transition. We have used experimental $\log ft^-=5.6$ for this transition and evaluated the corresponding results by treating it as $\log ft$ value for the allowed transition. With increase in temperature, the $7/2^-_1$ state at 154 keV in $^{33}$Mg has considerable occupancy after 0.5 GK but the $^{33}$Mg($7/2^-_{1}$) $\rightarrow$ $^{33}$Al($5/2^+_{\rm g.s.}$) is also first forbidden transition which leads to the decrease in $L_{34}$ after 0.5 GK. In addition, the neutrino luminosity is also evaluated for this Urca pair, which is shown in Fig. \ref{lum_33al}. In this case, the values of the mass fraction considered are $X_{\rm SB} = 4.0\times 10^{-6}$, $X_{\rm S} = 8.8\times 10^{-5}$, and $X_{\rm XRB} = 4.7\times 10^{-3}$ \cite{meisel2017}.

Now, we compare our calculated luminosities with those of Ref. \cite{schatz2014}, where the values for $T_9$ =0.51 and $X(A)$=1 are shown in Table 1 in units of 10$^{36}$ erg s$^{-1}$.
Their values are 24, 8.8 and 8.3 for the ($^{29}$Mg, $^{29}$Na), ($^{31}$Al, $^{31}$Mg) and ($^{33}$Al, $^{33}$Mg) pairs, respectively. 
The corresponding values for the luminosity obtained here by shell-model calculations with the CEFT, DJ16A, USDB and SDPF-MU are 11-15, 8-12, and 1.2 for the ($^{29}$Mg, $^{29}$Na), ($^{31}$Al, $^{31}$Mg) and ($^{33}$Al, $^{33}$Mg) pairs, respectively.
The value for the EEdf1 in the ($^{31}$Al, $^{31}$Mg) pair is 0.8, one order of magnitude smaller than the values of other interactions.  
For the ($^{29}$Mg, $^{29}$Na) pair, our shell-model value is about half of the QRPA value of Ref. \cite{schatz2014}, while for the ($^{31}$Al, $^{31}$Mg) pair our value is close to that of Ref. \cite{schatz2014} except for EEdf1. The results with the EEdf1 are close to those of the shell-model calculation of Ref. \cite{wang2021}. The difference in the cooling strength comes from the difference in the GT strength for the transition, $^{31}$Mg (3/2$^{+}$, 0.050 MeV) $\leftrightarrow$ $^{31}$Al (5/2$^{+}$, g.s.).
In case of the ($^{33}$Al, $^{33}$Mg) pair, in which the transitions between the ground states are first-forbidden, our shell-model calculation predicts a small value for the luminosity compared to that of \cite{schatz2014}.

\section{Summary and Conclusions} \label{Conclusion}

In this work, the GT strengths and weak rates that includes electron capture and $\beta^{-}$-decay rates in stellar environments for nuclear pairs with $A$ = 23, 25, 29, 31 and 33 are investigated by shell-model calculations with the use of {\it ab initio} $sd$-shell interactions as well as phenomenological interactions.
The Urca pairs, ($^{23}$Na, $^{23}$Ne), ($^{25}$Mg, $^{25}$Na) and ($^{25}$Na, $^{25}$Ne), are important for the cooling of O-Ne-Mg cores of eight to ten solar mass stars \cite{Toki2013, schwab2017}.
For the ($^{23}$Na, $^{23}$Ne) pair, it was shown that the Urca density can be assigned for {\it ab initio} interactions, CEFT and DJ16A, similar to the case of the phenomenological interaction, USDB, regardless of some difference in the calculated GT strength among the interactions. 
When available experimental data are taken into account, the weak rates for the ($^{23}$Na, $^{23}$Ne), ($^{25}$Mg, $^{25}$Na) and ($^{25}$Na, $^{25}$Ne) pairs become almost independent of the interactions and the Urca density that is common to the interactions can be assigned.  
In case of the ($^{25}$Na, $^{25}$Ne) pair, while transitions between the ground states are second forbidden, allowed GT transitions are possible between low-lying states with excitation energies below 100 keV. In such a case, an assignment of Urca density is possible at high temperatures. 
The nuclear Urca process by the ($^{25}$Na, $^{25}$Ne) pair is triggered in late heating stages of the evolution of the O-Ne-Mg core, when the heating by double electron captures on $^{20}$Ne occurs \cite{schwab2017, suzuki2022, kirsebom2019a, kirsebom2019b}.

The GT strengths and weak rates are also evaluated for the ($^{29}$Mg, $^{29}$Na), ($^{31}$Al, $^{31}$Mg), and ($^{33}$Al, $^{33}$Mg) nuclear pairs, which are important for the cooling of the neutron star crusts \cite{schatz2014}, as well as the ($^{31}$P, $^{31}$Si) pair, which is expected to play a role in the cooling and neutralization of the core in oxygen and silicon burnings in massive stars. For the nuclear pairs with  $A$ = 31 and 33, a phenomenological interaction in the $sd$-$pf$ shell, SDPF-MU, is used in addition to the effective interactions in the $sd$-shell, CEFT, DJ16A, and USDB. 
In case of the pairs with $A$ = 29 and 31, when available experimental data are included, the weak rates become nearly independent of the interactions, and the Urca density common to them can be assigned.
For the ($^{31}$Al, $^{31}$Mg) pair, though the transitions between the ground states are second-forbidden, there exists a 3/2$^{+}$ state at 50 keV in $^{31}$Mg, which results in allowed GT transitions between $^{31}$Al (5/2$^{+}$, g.s.) and $^{31}$Mg (3/2$^{+}$, 0.050 MeV) states. This situation is similar to the case of the ($^{25}$Na, $^{25}$Ne) pair.
The EEdf1 interaction obtained by the EKK method is also employed to evaluate the weak rates for the ($^{31}$Al, $^{31}$Mg) pair. Though the calculated rates are small compared to other interactions owing to the smaller GT strength, the Urca density assigned proves to be the same as the other interactions.   
For the ($^{33}$Al, $^{33}$Mg) pair, the transitions between the ground states are first forbidden, for which the experimental transition strength is used, and the GT strengths between positive-parity states are evaluated with the SDPF-MU interaction.
Though the calculated weak rates are rather small, the Urca density is assigned. 

The neutrino luminosities of the Urca processes in neutron star crusts are estimated by evaluating the intrinsic cooling strength for the nuclear Urca pairs using the weak rates obtained by the shell-model calculations. The mass fractions of the parent nuclei evaluated by crust cooling models of the accreting neutron star \cite{meisel2017} are used.
The cooling is found to depend sensitively on the crust temperature for the ($^{29}$Mg, $^{29}$Na) and ($^{31}$Al, $^{31}$Mg) pairs because of important contributions from low-lying excited states in the nuclei. The cooling effects of the present shell-model results of most of the interactions considered here are comparable to those of Ref. \cite{schatz2014} except for the case of the ($^{33}$Al, $^{33}$Mg) pair, in which the transitions between the ground states are first-forbidden.  As for the ($^{31}$Mg, $^{31}$Al) pair, the neutrino luminosity is sensitive to the transition strength between the $^{31}$Mg (3/2$^{+}$, 0.050 MeV) and $^{31}$Al (5/2$^{+}$, g.s.) states.

When available experimental excitation energies and GT strengths are taken into account in shell-model calculations, both {\it ab initio} and phenomenological interactions in the $sd$ shell, CEFT, DJ16A, and USDB, are found to lead to almost equivalent weak rates. This is mainly because the most important contributions often come from transitions between the ground states, for which experimental information is usually available. As for the intrinsic cooling strength and the neutrino luminosity of the Urca process in neutron star crusts, contributions from excited states become important at high temperatures, which results in variations among the interactions. The weak rates for the nuclei with mass number $A =17-28$ obtained by the USDB were tabulated in Ref. \cite {suzuki2016} to contribute to astrophysical studies. Here, we extend their work with the USDB interaction to the nuclei with $A =29-39$ for further astrophysical applications: we give the tabulated weak rates for the nuclei with $A =29-39$ obtained with the USDB including available experimental data and the screening effects, along with neutrino energy loss rates and $\gamma$-ray heating rates \cite{suppl}. Besides them, for the ($^{29}$Mg, $^{29}$Na), ($^{31}$Al, $^{31}$Mg) and ($^{33}$Al, $^{33}$Mg) pairs, the weak rates obtained with the SDPF-MU interaction are also tabulated \cite{suppl}.
We hope that the compiled data for the weak rates are helpful for the studies of the evolution of stars, supernovas and nucleosynthesis.

\vspace{-2mm}

 \section*{Acknowledgements}
This work is supported by a research grant from SERB (India), Grant No. CRG/2022/005167. S. S. would like to thank CSIR-HRDG (India) for the financial support for her Ph.D. thesis work. We acknowledge the National Supercomputing Mission (NSM) for providing computing resources of ‘PARAM Ganga’ at the Indian Institute of Technology Roorkee.

\end{document}